\documentclass[useAMS,usegraphicx]{mn2e}
\usepackage{amssymb}
\begin{document}

\title[H$_2$ from massive YSOs with methanol masers]{Testing the
circumstellar disk hypothesis:\ A search for H$_2$ outflow signatures from massive
YSOs with linearly distributed methanol masers}
\author[J. M. De Buizer]{J. M. De Buizer\thanks{
E-mail: jdebuizer@ctio.noao.edu}\\
Cerro Tololo Inter-American Observatory, Casilla 603, La Serena, Chile\thanks{Cerro Tololo Inter-American Observatory (CTIO) is operated by AURA, Inc. under
contract to the National Science Foundation.}}
\date{Accepted 2002 December 19. Received 2002 December 19; in original form
2002 November 22}
\maketitle

\begin{abstract}
The results of a survey searching for
outflows using near-infrared imaging is presented. Targets were chosen from a compiled list of
massive young stellar objects (YSOs) associated with methanol masers in linear
distributions. Presently, it is a widely held belief that these methanol
masers are found in (and delineate) circumstellar accretion disks around
massive stars. If this scenario is correct, one way to test the disk hypothesis is to
search for outflows perpendicular to the methanol maser distributions.
The main objective of the survey was to obtain wide-field
near-infrared images of the sites of linearly distributed methanol masers
using a narrow-band 2.12 $\micron$ filter. This filter is centered on the H$%
_{2}$ $v$=1--0 S(1) line; a shock diagnostic that has been shown to
successfully trace CO outflows from young stellar objects. 
Twenty-eight sources in total were imaged of which eighteen
sources display H$_2$ emission. Of these, only \textit{two} sources
showed emission found to be dominantly perpendicular to the methanol
maser distribution. Surprisingly, the H$_2$ emission in these fields is not distributed randomly, but instead the majority of sources are
found to have H$_{2}$ emission dominantly \textit{parallel} to their
distribution of methanol masers. These results seriously question the
hypothesis that methanol masers exist in circumstellar disks. The possibility that linearly distributed methanol masers are instead directly associated with outflows is discussed.  
\end{abstract}

\pagerange{\pageref{firstpage}--\pageref{lastpage}} 
\pubyear{2002}

\label{firstpage}


\section{Introduction}

Our understanding of star formation, despite decades of research, is still
quite limited. While the prescription for low mass
star formation is based on the ideas of creation via accretion (Shu,
Adams \& Lizano 1987), it is unknown if massive stars form in
this way. There are problems when applying the standard model of accretion
to the highest mass stars -- most notably the effects of radiation pressure which may inhibit any
further accretion once the star has accreted $\gtrsim $10 M$_{\sun}$. Since stars more massive than 10 M$_{\sun}$ do exist, and
since they tend to form in the middle of dense clusters, the idea of massive
stars forming through a process of coalescence of low mass stars or
protostars has been proposed (Bonnell, Bate \& Zinnecker 1998).
However, recent modelling by McKee \&
Tan (2002) has shown that despite all the alleged problems, the highest mass
stars can indeed be formed theoretically via accretion alone.

If massive stars do form via accretion, then at the least, in the earliest
stages of development they too must have accretion disks. So one way of
proving or disproving that massive stars form via accretion would be to try
to directly image these accretion disks around the most massive B\ and O
type stars. Despite many attempts at several different wavelengths,
there exists today no directly imaged accretion disk confirmed to exist
around a star of spectral type B2 or earlier. There are two main reasons why
it is more difficult to observe these accretion disks around massive stars
in comparison to low mass stars. First, regions of massive
star formation lie at distances of typically a few to 10 kpc away. This is much more
distant than regions of low mass star formation which are on the order of
100s of parsecs away. Second, the earliest stages of massive star formation
are difficult to observe. Massive stars form so quickly that they are still
accreting when they enter the ZAMS and are therefore enshrouded in their
placental cloud. The extinction towards a still-accreting massive star is so
high that no ultraviolet\ or optical light from the young star
will make it to the observer, and even near-infrared\ photons (1-2 $\micron$%
) are typically believed to suffer this extinction. Furthermore, because these
regions are so distant, resolution has been a problem in the mid- and
far-infrared. Interferometry in the millimeter and radio can achieve the
necessary resolution, but these wavelengths are not best suited for
observing thermal dust emission from accretion disks due to possible 
contamination by free-free radiation. The final alternative is the submillimeter, which can probe the cool thermal emission from a dust disk with minimal contamination. But we are still 5-10 years away from the submillimeter interferometric arrays that are necessary for achieving the resolution needed to resolve these disks.

We may be extremely limited when trying to observe these disks in their
thermal emission, however there is a non-thermal phenomenon that is thought
to trace these disks. Methanol masers tend to be distributed in the sky in
linear structures, and often with velocity gradients along the masers
which may indicate rotation (Norris et al. 1993). They are also believed to be signposts for massive
star formation. There is a growing belief that these linearly distributed
methanol masers exist in, and delineate, circumstellar disks around
massive stars. 

But do methanol masers really trace disks? More proof is needed than a line of masers displaying (perhaps) a rotating motion. Since several of these linear
distributions of methanol masers are of the order of 0\farcs5 -- 1\farcs5 in
size, and if they are indeed residing in disks, then perhaps one could in fact directly observe them. In the mid-infrared, resolutions of 0\farcs6 on a 4-m
class telescope and 0\farcs25 on a 10-m class telescope are achievable.
However, despite attempts to directly image these disks in the mid-infrared, results have been
inconclusive. The survey of De Buizer, Pina \& Telesco (2000) included ten
sources of linearly distributed methanol masers, but only three had
mid-infrared emission that was resolved using a 4-m telescope. All three
resolved sources were elongated in their thermal dust emission at the same
position angles as their methanol maser distributions. They argued, based on
several pieces of observational evidence, that their results were
consistent with the circumstellar disk hypothesis for methanol masers.

One of these three elongated mid-infrared sources was observed on a 3.6-m
telescope by Stecklum \& Kaufl (1998). They also argued that the elongated
source that they observed in the mid-infrared was a circumstellar disk around a
high mass star. However, follow-up observations by De\ Buizer et al. (2002a) using the
Keck 10-m, discovered that this elongated ``disk'' was in reality three
aligned mid-infrared sources. Consequently, direct detection of these disks is still a
problem in the mid-infrared from the standpoint of resolution. Furthermore,
it has been argued that in the mid-infrared it is difficult to tell if you
are observing dust emission from a circumstellar accretion disk or dust
emission from the placental envelope (Vinkovi\'{c} et al. 2000).
Corroborative evidence that linearly distributed methanol masers exist in
circumstellar disks, therefore, needs to come from something other than
direct detection of the accretion disks.

Fortunately, there is an \textit{indirect} way of testing whether or not
linearly distributed methanol masers exist in accretion disks. According to
the standard model of accretion, during the phase of stellar formation where
the star is being fed by an accretion disk, it is also
undergoing mass loss through a bipolar outflow. This bipolar outflow is
perpendicular to the plane of the accretion disk, and along the axis of
rotation. Therefore, one can search these sources of linearly distributed
methanol masers for evidence of outflow perpendicular to the methanol maser
position angle. Such evidence would create an extremely solid case for the hypothesis that these methanol masers exist in circumstellar
disks, without the need for their direct detection. Also, if collimated
outflows are found to be a general property of young massive stars, this
would be clear observational support for the idea that massive stars, like
low mass stars, do indeed form via accretion.

In this vein, a survey of these sources of linearly distributed methanol
masers was undertaken to search for signs of outflow. The main objective of
the survey was to image each site in the near-infrared with a narrowband
2.12 $\micron$ filter, which is centered on the H$_{2}$ $v$=1--0 S(1)
line. H$_{2}$ can be excited by collisions in shocks (i.e. McKee, Chernoff \& Hollenbach 1982), and specifically shocks associated with outflows from young stellar sources. Davis \& Eisl\"{o}ffel (1995) convincingly showed that H$_2$ emission traces shocks in CO outflows from low-mass YSOs. However, the excitation of H$_{2}$ emission can be due to another process, namely radiative excitation by UV photons (i.e. Burton 1992). By looking at the morphology of the H$_2$ emission one can, in principal, differentiate between what is most likely H$_2$ emission excited by UV fluorescence and what is likely H$_2$ excited by shocks. Since massive stars generally produce copious UV photons, thus forming ultracompact H{\scriptsize II} (UC H{\scriptsize II}) regions, one would expect radiatively excited H$_{2}$ to be present in the very near stellar environment of massive stars. On the other hand, H$_2$ emission from shocks in outflows is expected to exist in knots or regions offset from the central stellar engine. Therefore, by imaging these regions in the near-infrared one can look for structures associated with hydrogen
emission emanating from the locations of the methanol masers. Near-infrared observations of H$_2$ have
become a standard technique for observing molecular outflows, and it is a suitable initial step in testing the circumstellar disk hypothesis of methanol masers through outflow observations.

\section{Observations and Data Reduction}

The target list consisted of 28 maser sites compiled mostly from the articles
by Walsh et al. (1998), Phillips et al. (1998), and Norris et al. (1998). The coordinates for all of
these sites are shown in the Table 1. These sites all contain sources of
linear methanol maser distributions, many with velocity gradients along
their distributions indicative of rotation, and therefore represent the best
circumstellar disk candidates.

Observations of all sources were made in 2002 June on the CTIO\ Blanco 4-m
telescope in Chile using OSIRIS, the Ohio State InfraRed
Imager/Spectrometer. OSIRIS operates at wavelengths from 0.9 to 2.4 $\micron$
and uses a 1024$\times $1024 HAWAII HgCdTe array supplied by CTIO. All
observations were taken using the f/2.8 imaging mode, yielding a pixel scale
of 0\farcs403 pixel$^{-1}$, for a field of view of 233$\arcsec\times 233\arcsec$. Each source
was observed through 2 filters: the narrowband H$_{2}$ ($\lambda _{o}$=2.12 $%
\micron$, $\Delta \lambda $=0.027 $\micron$) filter centered on the $v$=1--0
S(1) line of H$_{2}$, and the narrowband $\lambda _{o}$=2.14 $\micron$ ($%
\Delta \lambda $=0.021 $\micron$) continuum-only filter.

Observations were performed by using a 9 element dither pattern with 20$%
\arcsec$ offsets. This pattern was performed three times for each source.
First time through the pattern, the H$_{2}$ filter was used with an exposure
time of 20 seconds in each of the 9 positions, and the pattern was repeated
once more with the same filter and exposure time. The third time through the
pattern, the filter was changed to the 2.14 $\micron$ continuum filter and
an exposure time of 20 seconds was again used in each element of the pattern.

The 9 images from each element in the dither pattern were then stacked and
median averaged to produce a sky frame where the source and background stars
were completely removed. Approximately half the sources observed were in
rather crowded areas of the Galactic plane or in areas of extended near-infrared
emission. In a few cases, extended emission was seen in the preliminary
reductions at the telescope, and the telescope was slewed 6$\arcmin$ north,
and a 9 element dither pattern was done on a less-crowded but nearby piece
of sky in both filters. In all other cases, frames from the source observed
just before or after were used in conjunction with the 9 images for the
source at hand. Once medianed, the sky frames were very clean. In this manner
a clean sky frame for both the H$_{2}$ and continuum filter were produced.
These sky frames were then subtracted from each image of the 9 element
dither pattern, and then the 9 sky-subtracted images were registered and
aligned to produce the final mosaicked image. The alignment of each element
was done with a chi-square algorithm to ensure accurate registration.

For each source this process yielded two H$_{2}$ images, and one continuum
image, all with exposure times that are effectively 3 minutes, except on
the edges of the images where there was less overlap of the dither frames,
and correspondingly a shorter effective exposure time. However, the edges of
these images were far enough away from the central source and were cropped
so that only the central portions of the images (where the collective exposure times
are 3 minutes) were used. Any of these images found not to have a zero mean
background or a gradient in the background had this residual sky emission
subtracted off by use of an algorithm that subtracts off a two-dimensional
polynomial surface of second order in x and y. Bright stars were found in
one of the H$_{2}$ images and the brightness of these sources were
determined using aperture photometry. These same stars were found on the
continuum image and again their brightnesses were found. Using the ratio in
brightness between the H$_{2}$ and continuum sources, the images were then
normalized so that these bright continuum-only stars on the frame were the
same brightness in both the H$_{2}$ and continuum image. The continuum image
was then shifted to match the positioning of the stars on the H$_{2}$ image,
again using a chi-square technique. The continuum image was then subtracted
from the H$_{2}$ image. This continuum subtraction technique was repeated
for the other H$_{2}$ image, and then the two continuum subtracted H$_{2}$
frames were stacked. This final image was then examined for H$_{2}$-only
emission. In most cases, once the continuum and H$_{2}$ images were
normalized and placed side-by-side, it was easy to differentiate between H$%
_{2}$-only or H$_{2}$-dominated sources and continuum-only sources on the
frames.

These final images were then compared to fields from the Digitized Sky
Survey (DSS) provided by the Space Science Telescope Institute (STScI).
Astrometric calibration of each field was found in this way, and the
accuracy of the DSS coordinates was checked by identifying Tycho-2 Catalog
stars in the fields. Once the images were in an absolute coordinate system,
the position of the methanol masers (known to an absolute astrometric
accuracy of $<$1$\arcsec$) were identified. Taking into account the
fact that the pixel size is 1\farcs7 for the DSS STScI images, and the
effects of seeing and PSF size, the overall accuracy of this method of
astrometry is estimated to be accurate to 3\farcs0. The error on the
position given by the DSS STScI images could be of the order of 4$\arcsec$
on the plate edges, but this large of an offset was never found in the
comparisons with the Tycho-2 sources on the fields.

No attempts were made to accurately flux calibrate the images in
either filter. One or two standard stars were observed throughout
the course of each night so a rough estimate of fluxes could be
determined if necessary. The main objective of this program was to
see if there is any H$_{2}$-only emission emanating from these
sources and if they are preferentially found to be collimated and
perpendicular to the methanol maser distributions. Therefore no
source fluxes will be included in the results presented here.

\section{Results}

Of the 28 maser targets observed, H$_{2}$ emission was detected from 18
sites (64\%). The distribution of the H$_{2}$ emission from these sites
takes on three forms: 1) extended diffuse areas of H$_{2}$-dominated
emission; 2) individual knots or blobs of H$_{2}$-only emission that range
in number, sometimes cometary-shaped; and 3) some combination of extended H$%
_{2}$-dominated emission with knots of H$_{2}$-only sources.

All of the methanol maser sites observed are shown in Figures 1 to 26. Each
figure displays a panel showing the H$_{2}$+continuum image of the region
around the masers, as well as the continuum-only image normalized to the same
intensity. If there is detectable H$_{2}$
emission in the region, then the figure also contains a residual H$_{2}$
image produced from the differencing of the continuum image and the H$_{2}$%
+continuum image. In many cases, the H$_{2}$-only \ and H$_{2}$-dominated
sources are so obvious that they can be discerned by eye when comparing the H%
$_{2}$+continuum image with the continuum images in each figure. However, in
some cases the H$_{2}$ emission can only be seen in the residual frame.

There is also ``noise'' on the residual H$_{2}$ frames because of the inability
to subtract off stellar continuum sources completely. This is due to changes in
seeing in the time between taking the H$_{2}$+continuum image and the continuum-only image, as
well as and diffraction spikes and internal reflections, especially for the brightest
sources. The differencing method produces a positive residual for any
emission seen exclusively in \textit{either} the H$_{2}$+continuum or
continuum-only image. Careful inspection was made to ensure
that noise sources were not mistaken for H$_{2}$ sources. An example of a diffraction spike can be seen in the upper left corner of the H$_2$+continuum and residual images in Figure 1. Examples of internal reflections can best be seen in the continuum image of Figure 17 as the ``double halos'' just north of the four bright continuum sources on the field. 

The position of the maser distributions is given by the crosses in each figure. One axis of
the cross was made longer than the other and rotated so that the long axis
indicates the position angle of the linear distribution of methanol masers.
Dashed circles were added to show the locations of the H$_{2}$ sources, and
ellipses encompass multiple sources and/or regions of H$_{2}$ emission. Again,
other residuals may be seen on the H$_{2}$ residual frame and are not
circled because they are confirmed to be noise. In order to check if the
majority of H$_{2}$ emission in the fields was predominantly perpendicular or
parallel, dashed lines were added to the residual H$%
_{2} $ frames of each figure where H$_{2}$ was detected. These dashed lines
divided each image into parallel and perpendicular quadrants, which are
marked. A target was deemed to have H$_{2}$ emission ``perpendicular'' if the
majority of the H$_{2}$ sources were found in the perpendicular quadrants.

Again, H$_2$ emission can be due to either UV fluorescence or shocks. One way of differentiating between shock and radiatively excited H$_{2}$ emission is to compare the line intensities of the 1--0 S(1) line of H$_{2}$ at 2.12 $\micron$ to the 2--1 S(1) line of H$_{2}$ at 2.25 $\micron$. However, since targets were not imaged through a 2--1 S(1) H$_{2}$ filter, the only way to determine the likely origin of the H$_{2}$ emission is by looking at the morphology of the H$_{2}$ emission. For some targets in the survey, extended star-forming clouds were present,
and H$_{2}$ emission from these regions is most likely dominated by UV\
fluorescence rather than shock excited by outflows. These types of sources are deemed ``not outflow'' in nature. Table 1 lists each source, with the status of the H$_{2}$ emission for each as ``parallel'',
``perpendicular'', ``not outflow'', or ``confused''. A site is labelled ``confused'' if it can not be fit into any one of the other three categories. The reasons for calling a maser site confused are site-specific, so refer to \S3.1 on individual sources for clarification. 

Contrary to what was expected, of the 18 sites with H$_{2}$ emission, only
\textit{two} sites were found to have H$_{2}$ emission dominantly
perpendicular to their methanol maser position angles. The majority of the H$_{2}$ sources, are found to be
\textit{parallel} to the maser distributions. Of the 18 sources where H$_{2}$
was detected, 12 have H$_{2}$ distributed predominantly in quadrants
parallel to the maser position angle.

\subsection{Individual sources}

Below is a summary of what is known about each source, as it pertains to
the interpretation of the H$_{2}$ emission on each field. There are several
sources that have complex fields, and such interpretation is difficult.
This section is intended as a complete guide to the figures and Table 1.
Though most of the sources in the survey are from Walsh et al. (1998) who
used IRAS\ names, throughout this paper the names are given in terms of
their galactic coordinates. This type of naming convention is more
informative and also helps resolve confusion if there is more than one maser
group associated with a particular IRAS\ source. For completeness, however,
the IRAS\ names will be given in the section headings below and in Table 1.

\begin{table*}
\begin{minipage}{152mm}
\caption{List of targets observed in this survey. The first column is the name of the target in galactic coordinates and the second column is the IRAS name. The third and fourth columns give the source coordinates. The fifth column gives the methanol maser distribution position angle. The sixth column describes the H$_2$ emission found in the target field. The seventh and eighth columns list if there is radio continuum or near-infrared emission directly coincident with the maser location.}
\begin{tabular}{lccccccc}
\hline
Target & IRAS Name & Right Ascension & Declination & Maser & H$_2$ & \multicolumn{2}{c}{Maser association} \\
& & (J2000) & (J2000) & p.a. &  & Radio? & NIR?\\ 
\hline
G305.21+0.21   & 13079-6218 & 13 11 13.72 & -62 34 41.6 & 25\degr & parallel?    & N$^p$ & N\\
G308.918+0.123 & 13395-6153 & 13 43 01.75 & -62 08 51.3 & 137\degr& parallel     & Y$^p$ & Y\\
G309.92+0.48   & 13471-6120 & 13 50 41.76 & -61 35 10.1 & 30\degr & no detection & Y$^p$ & Y\\
G312.11+0.26   & 14050-6056 & 14 08 49.30 & -61 13 26.0 & 166\degr& no detection & N$^w$ & N\\
G313.77-0.86   & 14212-6131 & 14 25 01.62 & -61 44 58.1 & 135\degr& parallel     & N$^w$ & Y\\
G316.81-0.06   & 14416-5937 & 14 45 26.44 & -59 49 16.4 & 1\degr  & not outflow  & N$^w$ & Y\\
G318.95-0.20   &            & 15 00 55.40 & -58 58 53.0 & 151\degr& parallel     & N$^e$ & Y\\
G320.23-0.28   & 15061-5814 & 15 09 51.95 & -58 25 38.1 & 86\degr & parallel     & N$^w$ & N\\
G321.031-0.484 & 15122-5801 & 15 15 51.64 & -58 11 17.4 & 0\degr  & parallel?    & N$^w$ & Y\\
G321.034-0.483 & 15122-5801 & 15 15 52.52 & -58 11 07.2 & 85\degr & parallel?    & N$^w$ & Y\\
G327.120+0.511 & 15437-5343 & 15 47 32.71 & -53 52 38.5 & 150\degr& no detection & Y$^p$ & Y\\
G327.402+0.445 & 15454-5335 & 15 49 19.50 & -53 45 13.9 & 62\degr & no detection & Y$^p$ & Y\\
G328.81+0.63   & 15520-5234 & 15 55 48.61 & -52 43 06.2 & 86\degr & parallel     & Y$^w$ & Y\\
G331.132-0.244 & 16071-5142 & 16 10 59.74 & -51 50 22.7 & 90\degr & parallel     & Y$^p$ & N\\
G331.28-0.19   & 16076-5134 & 16 11 26.60 & -51 41 56.6 & 166\degr& perpendicular& Y$^p$ & N\\
G335.789+0.174 &            & 16 29 47.33 & -48 15 52.4 & 136\degr& parallel?    & N$^p$ & Y\\
G336.43-0.26   & 16303-4758 & 16 34 20.34 & -48 05 32.5 & 163\degr& no detection & N$^p$ & Y\\
G337.705-0.053 & 16348-4654 & 16 38 29.61 & -47 00 35.7 & 137\degr& no detection & Y$^p$ & Y\\
G339.88-1.26   & 16484-4603 & 16 52 04.66 & -46 08 34.2 & 137\degr& ?            & Y$^e$ & Y\\
G339.95-0.54   & 16455-4531 & 16 49 07.99 & -45 37 58.5 & 122\degr& no detection & N$^w$ & N\\
G344.23-0.57   & 17006-4215 & 17 04 07.70 & -42 18 39.1 & 117\degr& no detection & N$^w$ & N\\
G345.01+1.79   & 16533-4009 & 16 56 47.56 & -40 14 26.2 & 78\degr & parallel?    & Y$^w$ & N\\
G345.01+1.80   & 16533-4009 & 16 56 46.80 & -40 14 09.1 & 30\degr & parallel?    & N$^w$ & N\\
G348.71-1.04   & 17167-3854 & 17 20 04.02 & -38 58 30.0 & 152\degr& not outflow  & N$^w$ & N\\
G353.410-0.360 & 17271-3439 & 17 30 26.17 & -34 41 45.6 & 153\degr& not outflow  & Y$^p$ & N\\
G00.70-0.04    & 17441-2822 & 17 47 24.74 & -28 21 43.7 & 51\degr & no detection & N$^w$ & Y\\
G10.47+0.03    & 18056-1952 & 18 08 38.21 & -19 51 49.6 & 98\degr & no detection & Y$^w$ & N\\
G11.50-1.49    & 18134-1942 & 18 16 22.13 & -19 41 27.3 & 174\degr& perpendicular& N$^w$ & N\\
\hline
\end{tabular}
\medskip \\
? denotes that there is confusion associated with the H$_2$ observations. See \S3.1 for details.\\
{\em $^p$} denotes that the result is from Phillips et al. (1998).\\
{\em $^w$} denotes that the result is from Walsh et al. (1998).\\
{\em $^e$} denotes that the result is from Ellingsen, Norris \& McCulloch (1996).\\
\end{minipage}
\end{table*}

\subsubsection{G305.21+0.21 (IRAS 13079-6218)}

G305.21+0.21 is a linear distribution of 4 bright maser spots as seen by Norris
et al. (1993), with 3 weaker spots found by Phillips et al. (1998).
These seven maser spots are not distributed in a straight line, but more like in
a flattened structure with a position angle of $\thicksim $25$\degr$. The
observed field also contains the methanol maser site known as
G305.20+0.21, which contains two
methanol maser spots (Norris et al. 1993) in a tight grouping. Again, other
weaker spots were found here by Phillips et al. (1998). These two
maser sites are only separated by $\thicksim $22$\arcsec$. Their close
proximity to the H$_{2}$ emission detected here leads
to uncertainty in which maser site the H$_{2}$ emission is associated with, or
if the H$_{2}$ emission is being produced from both sites. There are two H$%
_{2}$-only sources, one located $\thicksim $10$\arcsec$ south, and the other
$\thicksim $37$\arcsec$ southwest, of G305.21+0.21. There is also an
extended H$_{2}$-dominated source located $\thicksim $22$\arcsec$
southeast of G302.21+0.21 (Figure 1). If G305.21+0.21 is solely responsible
for all the H$_{2}$ emission (which may be the case since it appears closer
to all three H$_{2}$ sources) then the H$_{2}$ sources are distributed more
parallel than perpendicular to the methanol maser distribution. However,
because these H$_{2}$ sources could also be generated in part or solely by
the source of the G305.20+0.21 masers, this H$_{2}$ distribution for this site is labled
``parallel but confused'' in Table 1.

There is a very bright near-infrared source coincident with G305.20+0.21 that is surrounded by extended emission
also seen by Walsh et al. (1999) who determine its magnitude to be $M_{K}$ =
7.0, and $M_{L}$ = 3.9. There is a compact and extremely bright
mid-infrared source at this maser location as well (F$_{10\micron}$ = $\thicksim $29 Jy; De Buizer et al. 2000). G305.21+0.21 was not found to have near-infrared emission by
Walsh et al. (1999), however in the data presented here, there is indeed a
small, compact near-infrared continuum source coincident with the maser
position. A rough estimate of its apparent magnitude at 2.14 $\micron$ is $%
M_{2.14\micron}$ = 13.6. However, there is no color information in this data
set, therefore this source may be a field star. Neither of the maser
positions are associated with radio continuum emission, however, the
extended near-infrared emission containing G305.20+0.21 appears to be a
large (15$\arcsec$$\times$10$\arcsec$) H{\scriptsize II} region (Phillips et al. 1998).
This H{\scriptsize II} region can also bee seen at mid-infrared wavelengths (De Buizer et al. 2000; Walsh et al. 2001).

\subsubsection{G308.918+0.123 (IRAS 13395-6153)}

This maser site contains the most H$_{2}$-only sources of any target
observed in this survey. All 15 of these sources can be seen in Figure 2. The line of
masers for this site is comprised of only 4 maser spots (Phillips et al.
1998). Surprisingly, there is no detectable extended H$_{2}$ emission. The
vast majority of the 15 H$_{2}$ sources (10) are distributed within a
region $\leq $45$\degr$ from parallel to the maser distribution angle of $%
\thicksim $137$\degr$.

Phillips et al. (1998) show that these masers are on the northern edge of an
apparently spherical, extended H{\scriptsize II} region that has a radius of $\thicksim$6$\arcsec$.
Given the astrometry in this work, it was found that the masers are coincident
with the peak of a bright near-infrared continuum source, but there does not
appear to be any large H{\scriptsize II} region here seen in reflected dust emission in the near-infrared.
Phillips et al. (1998) speculate that the radio continuum could be the ionized
component of an outflow from the central source, which may be consistent in angle
with these observations, however they caution that the narrow velocity range
of the radio continuum make it unlikely.

\subsubsection{G309.92+0.48 (IRAS 13471-6120)}

The distribution of masers at this site is more arc-like than linear, and
spans almost a full arcsecond. This site contains 10 maser spots (Phillips
et al. 1998) with a very well-defined velocity gradient along the spots
that is argued to be consistent with a systemic rotation. For this and
several other reasons, this site has been characterized as a good
circumstellar disk candidate by De Buizer et al. (2000) and De
Buizer (2001), who observed the source to be elongated in thermal dust
emission at the same position angle as the maser distribution.

However, contrary to what was expected, no H$_{2}$
emission was found perpendicular to this maser distribution (Figure 3). In
fact, no detectable H$_{2}$ emission was found in the field at all. This is
consistent with the observations of Oliva \& Moorwood (1986), who do not
detect any H$_{2}$ emission spectroscopically from this region within a 30$%
\arcsec$ aperture. Nonetheless, the masers are associated with a bright
near-infrared continuum source, that was also seen at mid-infrared
wavelengths by De Buizer et al. (2000) and Walsh et al. (2001),
and is also a radio strong radio continuum source (Phillips et al. 1998).

\subsubsection{G312.11+0.26 (IRAS 14050-6056)}

This site contains 4 bright maser spots in a line (Walsh et al. 1998), but they
are isolated and offset more than 3$\arcmin$ from the nearest UC H{\scriptsize II}\
region and more than 4$\arcmin$ from the position of IRAS\ 14050-6056.
The near-infrared continuum image in Figure 4 shows no signs of
extended near-infrared emission within a 40$\arcsec$ radius of the maser
position. Furthermore, there is no detectable near-infrared emission at the
location of the masers. Likewise, there is no sign of H$_{2}$ emission from
this source. Unpublished observations of this site in the mid-infrared 
(De Buizer, in preparation) have also failed to detect any
thermal dust emission at this location. This site, therefore, has none of
the normal characteristics of a star forming region. If a massive star is
forming here, it must be extremely young and forming in relative isolation,
which is contrary to the established idea that massive stars form in
clusters at the center of giant molecular clouds.

\subsubsection{G313.77-0.86 (IRAS 14212-6131)}

There is a hydroxyl maser (Caswell 1998) coincident in location to
the group of 4 methanol maser spots at this location. These 4 spots are
shown to be linearly distributed in Walsh et al. (1998), but due to the size
of the error bars in the maser positions, the reality of the linear
distribution of these spots is perhaps debatable.

In Figure 5, extended near-infrared emission can be seen in a structure elongated roughly east-west. The majority of H$_2$ emission here is associated with this diffuse near-infrared lobe of emission. Furthermore, except for one H$_2$ source to the southwest, all of the H$_2$ emission at this site is found in one quadrant parallel to the methanol maser position angle. Therefore, in Table 1 this source is listed as dominantly ``parallel.''

The methanol masers are coincident with a extended near-infrared source that contains no radio emission (Walsh et al. 1998). Interestingly, the near-infrared source appears to be cometary shaped. This can not be a cometary UCH{\scriptsize II} region, due to the lack of radio emission from this site. The near-infrared emission associated with the masers may appear to be extended simply because of environmental effects or due to several unresolved young stellar sources present at this location. However, an alternative scenario can be suggested based on the morphology of the whole region. In Figure 5, there is a group of three bright near-infrared sources at
the location ($\Delta \alpha $, $\Delta \delta )$ = (-37, +10). The opposite side of the elongated lobe of near-infrared emission ends near these bright sources. Assuming the near-infrared and H$_2$ emission is due to outflow, one could argue that instead of the maser location being the center of outflow lobe, the masers could be at the \textit{head} of the outflow lobe and one of these bright sources is at the center. This would explain the cometary or ``bow-shock'' shape of the near-infrared source associated with the masers. This is the only source in the survey where there is evidence pointing towards the methanol masers existing in the shock-excited area at the head of the outflow, rather than the center.

There is another cluster of 3 methanol masers 23$\arcsec$ east and off the
edge of the image presented in Figure 5 at about the same declination as the
three bright near-infrared sources.

\subsubsection{G316.81-0.06 (IRAS 14416-5937)}

The masers at this location are in the middle of a spectacular and large dust
cloud complex. The near-infrared images in Figure 6 show the expansive
region taken up by the near-infrared emission alone, with a dark lane of
dust cutting through the region diagonally. This region is full of extended
mid-infrared emission as well (Walsh et al. 2001), and some radio continuum
emission (Walsh et al 1998).

The H$_{2}$ emission at this site comes from many areas of the cloud. The
masers are situated on the edge of a what appears to be a ``ring'' of H$_{2}$%
-dominated emission. Due to the fact that the masers reside in a star-forming cloud complex, it is more likely that the H$_2$ emission is produced by UV fluorescence than by shocks. Such rings of H$_2$ near massive YSOs has also been seen in IRAS 20293+3952 and IRAS 05358+3543 by Kumar, Bachiller \& Davis (2002). For this reason, this maser site is described as being ``not an outflow'' in Table 1.

\subsubsection{G318.95-0.20}

This site has 7 maser spots distributed in a linear fashion at an angle of $%
\thicksim$151$\degr$ (Norris et al. 1998). The masers are coincident with a
very bright near-infrared continuum source (Figure 7), which is also seen in
the mid-infrared (De Buizer et al. 2000; Walsh et al. 2001).
There does not appear to be any radio continuum from this source, however
(Ellingsen, Norris \& McCulloch 1996).

There are several sources of H$_{2}$-only and H$_{2}$-dominated emission
coming from this source, all of which are $\leq50\degr$ from parallel.
Interestingly, this same source was looked at in H$_{2}$ by Lee et al.
(2001), and apart from the lower level emission, the major sources that are
seen in Figure 7 can also be seen in the work of Lee et al. (2001). They
conclude that there is no definitive angle of emission, probably due to
confusion of the bright sources with the low-level emission. This low-level
emission can come from poor subtraction of the continuum from the H$_{2}$
image, and is a difficult problem to avoid. For this reason, the brighter H$%
_{2}$ sources are the most convincing and may have the greatest validity. Even
so, the majority of the H$_{2}$ emission in the image of Lee et al. (2001)
is still $\leq $45$\degr$ of being parallel to the maser distribution position
angle. While there is disagreement on the morphology of the H$_{2}$
emission, Lee et al. (2001) do argue that the H$_{2}$ emission is most
likely shock excited rather than due to UV fluorescence.

\subsubsection{G320.23-0.28 (IRAS 15061-5814)}

Of all the sources in the survey, this region contains H$_2$ emission that most closely resembles a bipolar outflow morphology. This maser site contains 10 maser spots, 9 of which are distributed into a tight
linear distribution spanning 0\farcs5 (Walsh et al. 1998). The tenth maser
spot lies 0\farcs2 north of the center of this distribution. Even including
this tenth maser spot, the distribution is clearly elongated at an angle of
about 86$\degr$. The H$_{2}$ emission in this region is distributed on either side of the
masers and extremely close to parallel (Figure 8). The H$_{2}$ emission to
the west is almost all H$_{2}$-only emission, whereas the emission to the
east is H$_{2}$-dominated and can best be seen in the residual H$_{2}$ frame
of Figure 8. Walsh et al. (1998) find a small radio continuum source 5$%
\arcsec$ north of the location of the eastern H$_{2}$ source. There does not
appear to be an association with any near-infrared source at this location.
This source not only represents the best case of an outflow from a methanol maser source in this survey, it is also the best candidate for further observations in disproving the circumstellar disk hypothesis for linearly distributed methanol masers.

\subsubsection{G321.031-0.484 \& G321.034-0.483 (IRAS 15122-5801)}

There are two sites of linearly distributed methanol masers in this
field separated by 10$\arcsec$ (Walsh et al. 1998). The northern maser site is G321.034-0.483
and has nine maser spots distributed in a line that is almost perpendicular to
the maser group in the south, G321.031-0.484, which is composed of 6 maser spots.
G321.031-0.484 is closest to all of the H$_{2}$ emission found here (Figure
9). Though an extended source of H$_{2}$ emission is located perpendicular
to the maser distribution of G321.031-0.484, the majority of the H$_{2}$
emission is found within the parallel quadrant. Of all the H$_{2}$ sources
on the field, the source that is perpendicular to G321.031-0.484 lies
closest to G321.031-0.484 and lies in its parallel quadrant. This extended H$%
_{2}$ source is also elongated radially to the position of G321.031-0.484, and
so may be associated. In this scenario then, the H$_{2}$ emission here would
be predominantly parallel to both maser sources. Regardless of these
arguments, the presence of both maser groups leads to some confusion as to
which source is responsible for what H$_{2}$ emission, if any. Therefore
both of these maser groups are listed as ``parallel but confused'' in Table
1.

\subsubsection{G327.120+0.511 (IRAS 15437-5343)}

This maser distribution is not only linear, but there is a well-defined
velocity gradient along the 5 maser spots (Phillips et al. 1998). The masers
appear to be coincident with a bright near-infrared source (Figure 10) that
also has faint radio continuum emission (Phillips et al. 1998). Interestingly, the extended near-infrared
continuum emission appears morphologically similar to a single-sided outflow
originating from the location of the masers. However,
there does not appear to be any H$_{2}$ emission in the area, and therefore is not likely an outflow. This is
consistent with the observations of Oliva \& Moorwood (1986), who do not
detect any H$_{2}$ emission spectroscopically from this region within a 30$%
\arcsec$ aperture. 

\subsubsection{G327.402+0.445 (IRAS 15454-5335)}

This site actually contains two sites of methanol masers, G327.402+0.444
and G327.402+0.445, as well as an isolated, single methanol maser spot
G327.402+0.444E (Phillips et al. 1998). However, all three maser sites lie within 5$\arcsec$ of each other. G327.402+0.445 has a linear maser
distribution consisting of 5 maser spots with a well-defined velocity gradient. This maser site is coincident with the center of a radio
continuum region seen by Phillips et al. (1998). G327.402+0.444 is also
claimed by Phillips et al. (1998) to be linearly distributed if one ignores
maser spot `B'. However if one instead ignores the weakest source, maser
spot `A' (or indeed any other maser spot in the group), then the
distribution would not appear linear at all. For this reason, this methanol
maser site is not considered to be linearly distributed in this paper.

In Figure 11, the masers of G327.402+0.445 are shown. G327.402+0.444 and
G327.402+0.444E lie ~2$\arcsec$ and ~5$\arcsec$, respectively, to the east and are not shown. All of the methanol masers here lie on the southern edge of a extended near-infrared
source, however there is no detectable H$_{2}$ in the region. Perhaps the
extended near-infrared emission that is present is related to the northern
part of the radio continuum region seen by Phillips et al. (1998).

\subsubsection{G328.81+0.63 (IRAS 15520-5234)}

This linear distribution of 9 methanol maser spots (Norris et al. 1998) is also
associated with a mid-infrared source (De Buizer et al. 2000) and
an UC H{\scriptsize II} region (Walsh et al. 1998). The masers lie $\thicksim $5$\arcsec$
south of a very bright near-infrared source (Figure 12), but according to
Osterloh, Henning \& Launhardt (1997) this is a reddened foreground star
that is not associated with the masers. The masers do appear to be
associated with near-infrared continuum emission in Figure 12. There are 4
sources of H$_{2}$-only emission near the masers, all located to the east of
the maser group and within $\thicksim $40$\degr$ of being parallel to the
position angle of the methanol masers.

\subsubsection{G331.132-0.244 (IRAS 16071-5142)}

The 9 methanol maser spots here are arranged predominantly at an angle of $%
\thicksim$90$\degr$, and have a velocity gradient along the spot
distribution (Phillips et al. 1998). An extended region of radio continuum
emission is found to be associated with the masers (Phillips et al. 1998,
Walsh et al. 1998), and the source appears to exist very close to (but not
coincident with) a bright near-infrared continuum source (Figure 13). There
is only one H$_{2}$-only source in the field, and it is extended and offset $\thicksim $8$%
\arcsec$ from the position of the masers at an angle close to parallel with
the maser distribution position angle.

\subsubsection{G331.28-0.19 (IRAS 16076-5134)}

This site contains eleven methanol maser spots, and though they are not in a tight
line, they are distributed in a flattened structure at an angle of $%
\thicksim $166$\degr$. Phillips et al. (1998) find extended radio
continuum emission enveloping this maser region and extending to the
southeast for more than 8$\arcsec$. It is possible that this could be the
partially ionized component to an outflow from the maser location, in which
case the outflow would be close to parallel to the maser distribution angle. The
other explanation for the radio continuum emission could be that is is an
extended UC H{\scriptsize II} region, however it is interesting to note that there is no
near-infrared emission coming from it (Figure 14), as is often seen in more
developed UC H{\scriptsize II} regions. This leans the evidence in favor of the radio
continuum coming from an outflow. However there is an extended, diffuse,
region of near-infrared continuum emission to the west of the location of
the masers that is not seen in the radio. In addition, this diffuse source
does indeed have a significant amount of H$_{2}$ associated with it. This H$%
_{2}$ emission, however, is positioned perpendicular to the methanol maser
position angle. Similar observations by Lee et al. (2001) of this region
show this same bright H$_{2}$ source, and it is argued by those authors as
being evidence of outflow and that the methanol masers do indeed arise for a circumstellar
disk. However, if this is indeed the outflow from the maser location, and
the radio emission is not a UC H{\scriptsize II} region, the nature of the radio continuum
is then a mystery.

\subsubsection{G335.789+0.174}

There are 11 maser spots at this sight, 6 of which are concentrated in a
linear distribution at a position angle of $\thicksim $136$\degr$ spanning
0\farcs1 (Phillips et al. 1998). The other masers are found in separate maser sites consisting of 3 and 2 spots
offset (0\farcs2 and 0\farcs3 away, respectively) from this main linear
distribution, and it is not clear if the masers are related or associated
with different sources. Observations in the near-infrared of this region
(Figure 15) show that the masers are located 20$\arcsec$ from a `ridge' of
extended emission running diagonally across the field. The masers are
coincident with a near-infrared continuum source, however because there is
no color information it can not be said with certainty if this near-infrared
source is the exciting source for the maser emission. The majority of the H$%
_{2}$ emission found here is associated with the edge of this ridge of
near-infrared emission which is most likely a photodissociation region. Therefore, the H$_2$ in this ridge is most likely excited by fluorescence, however there are a few components offset from the
ridge that lie parallel to the maser position angle (Figure 15). Though this
offset H$_{2}$ emission is parallel to the maser position angle, it is still
difficult to discern if the emission is, in fact, associated with the maser
source or part of the photodissociation ridge. Therefore the H$_{2}$ emission from this site is referred to as
`parallel but confused' in Table 1.

\subsubsection{G336.43-0.26 (IRAS\ 16306-4758)}

The methanol masers at this site were shown to have a linear distribution by
Norris et al. (1998), however the work of Phillips et al. (1998) added
additional components to the maser group, making it appear much less
linearly distributed. However, the rough, overall appearance of the maser
group is still elongated, though not overwhelmingly so. This maser site is
another interesting region like that of G312.11+0.26, where there appears to
be little evidence of star formation, although not as extreme. No radio
emission was detected here by Phillips et al. (1998), there was no detected
thermal dust continuum emission in the mid-infrared by De Buizer et al. (2000), and no OH\ masers were detected here by Caswell et al.
(1995). Furthermore, as seen in Figure 16, there is no evidence of H$_{2}$
emission. Walsh et al. (1999) also claim there is no near-infrared source
at the maser location. However, comparing their near-infrared images with
the ones presented here reveal that Walsh et al. (1999) erroneously placed
the masers 41$\arcsec$ southeast of the actual maser location in their
image. With the masers in the correct location in Figure 16, it can be seen
that there is some extended near-infrared continuum emission present near
the position of the methanol masers.

\subsubsection{G337.705-0.053 (IRAS\ 16348-4654)}

This site has 10 methanol maser spots linearly distributed with a velocity
gradient (Phillips et al. 1998). There is also another site consisting
of 2 methanol maser spots, G337.703-0.053, also nearby (Figure 17). The
site of linearly distributed masers is coincident with the peak of an
unresolved H{\scriptsize II} region seen by both Philips et al. (1998) and Walsh et al.
(1998). In the near-infrared the masers appear coincident (within the errors
of astrometry) with a continuum source as well. Whether this is the
near-infrared component of emission from the UC H{\scriptsize II}\ region or a field star
is not known. There is no detectable H$_{2}$ emission in the field.

\subsubsection{G339.88-1.26 (IRAS 16484-4603)}

The properties and observations of this site have been explored in detail by
De Buizer et al. (2002a). There are 49 methanol maser spots at this site in a
linear distribution and spread over 1\farcs5. There is extended radio
continuum emission present (Ellingsen et al. 1996), with the
methanol masers slightly offset to the south of the radio peak. There are
three mid-infrared sources coincident with the masers as well. De\ Buizer et
al. (2002a) speculate that the radio continuum, which is extended at a
position angle perpendicular to the maser distribution, may be the ionized
component of an outflow. Inspection of the near-infrared images of this
region (Figure 18) shows extended continuum emission in the direction
perpendicular to the maser distribution position angle. However, there is
also extended near-infrared continuum emission parallel to the masers as
well, seen by an outflow-shaped, elongated, and diffuse source that is
centered $\thicksim $60$\arcsec$ northwest of the maser location. The H$_{2}$
emission image does not help solve this dilemma, since there are two sources
of H$_{2}$ emission, with one located in a quadrant parallel to the maser
distribution and the other in a perpendicular quadrant (Figure 18). Due to the lack of H$_{2}$ present, the extended near-infrared emission here is most likely associated with the nearby star formation region and not an outflow.
Therefore, the nature of the H$_{2}$ sources, either shock excited or UV
excited, can not be ascertained either. The lack of copious amounts of H$%
_{2} $ emission could be viewed as being inconsistent with the speculation
that the radio emission here is delineating an outflow. One could argue that
the radio emission could be simply an elongated UC H{\scriptsize II} region, and the weaker
features speculated as being periodic mass ejection features may be
artificial by-products of the data reduction. However, further radio
continuum observations and observations of other outflow diagnostics, such
as HCO$^+$ could help solve this problem.

\subsubsection{G339.95-0.54 (IRAS 16455-4531)}

The linearly distributed methanol masers at this site are offset by $%
\thicksim $90$\arcsec$ from a UC H{\scriptsize II}\ region which has one methanol maser
spot associated with it (Walsh et al. 1998). This linearly distributed
group of masers has 16 maser spots. The near-infrared images of this
site show that the masers lie near a small region of low signal-to-noise
extended continuum emission, however there appears to be no clear evidence
of H$_{2}$ emission on the field (Figure 19). The extended emission region
is perpendicular to the maser distribution, however, and the masers are
slightly offset from the edge of this region where one would expect
morphologically an outflow to originate.

\subsubsection{G344.23-0.57 (IRAS 17006-4215)}

This site contains 10 methanol maser spots in a tight linear distribution (Walsh
et al. 1998) offset by $\thicksim $90$\arcsec$ from a UC H{\scriptsize II}\ region that
lies to the southeast. Walsh et al. (1999) detect no near-infrared component
at the maser location in \textit{K} or \textit{L} bands, consistent with the non-detection
presented here (Figure 20). Furthermore, there is no detectable H$_{2}$ in
the field.

\subsubsection{G345.01+1.79 \& G345.01+1.80 (IRAS 16533-4009)}

Separated by only 19$\arcsec$, G345.01+1.79 and G345.01+1.80 both display
methanol maser distributions that are linear. Norris et al. (1998) found
G345.01+1.79 to have 6 methanol maser spots distributed over $\thicksim $0\farcs%
3, whereas G345.01+1.80 has 14 methanol maser spots in a tight linear distribution
spanning the same angular size. Walsh et al. (1999) found both sites to be
lacking emission at \textit{K}, but G345.01+1.79 was found to have a
component in the \textit{L} band. Both sites have a mid-infrared component
at their maser locations (De Buizer et al. 2000; Walsh et al.
2001), but only G345.01+1.79 has radio continuum emission (Walsh et al.
1998). Due to the close spacing of the two methanol maser sources, it is
difficult to say for sure which maser source the H$_{2}$ emission in Figure
21 is associated with, or indeed if both are associated with the emission.
There is a long, extended region of near-infrared emission offset $\thicksim
$15$\arcsec$ from G345.01+1.79 and located exactly parallel to the methanol
maser distribution. It is also elongated along this direction. The H$_{2}$ emission
is dominantly from this elongated source, although there are two small H$%
_{2} $ sources $\thicksim $30$\arcsec$ north of G345.01+1.80. All of the H$%
_{2}$ emission on the field is $\lesssim $45$\degr$ from parallel with the
maser distribution of G345.01+1.80 (Figure 21). The two northern sources of H%
$_{2}$ are more likely to be associated with G345.01+1.80 than G345.01+1.79
because of proximity, and the rest of the H$_{2}$ emission in the field is
also $\leq $45$\degr$ from parallel to the maser distribution angle of
G345.01+1.79. Therefore, even though there is confusion regarding which
source is responsible for the H$_{2}$ emission, it can be said that the H$%
_{2}$ sources are consistent with being distributed more parallel than
perpendicular to \textit{both} G345.01+1.79 and G345.01+1.80. Therefore,
both sources are labelled ``parallel but confused'' in Table 1.

\subsubsection{G348.71-1.04 (IRAS 17167-3854)}

The 11 maser spots that make up this linear distribution (Walsh et al. 1998) are situated in the
middle of the impressively extended ($R_{cloud}$ = 4$\arcmin$) star
formation region RCW 122, also known as BFS 65 (Figure 22). Due to the fact that this
region contains widespread ionized gas and is extended in all directions
from the maser location, the diffuse and extended H$_{2}$ emission found
here (Figure 22) is most likely due to UV fluorescence. Therefore, in Table
1 this source is labelled ``not an outflow.'' Even though there is extended emission present, there is no near-infrared or radio component directly associated with the maser location.

\subsubsection{G353.410-0.360 (IRAS 17271-3439)}

The five methanol maser spots comprising this group are linearly distributed with
a well-defined velocity gradient along the spots (Phillips et al. 1998). The maser site lies $%
\thicksim $30$\arcsec$ from the middle of an extended near-infrared cloud of
emission (1\farcm5$\times 1\arcmin$), but is not coincident with a
near-infrared continuum source itself (Figure 23). The methanol masers are,
however, coincident with the location of OH\ masers and a slightly extended
radio continuum source (Forster \& Caswell 2000). The H$_{2}$ emission is
scattered weakly throughout the extended region of near-infrared emission,
and is brightest on the side of the region furthest from the masers. There
is also extended radio continuum emission in this area seen by Forster and
Caswell (2000), however, they do not image the radio continuum as far east
as the location of the brightest H$_{2}$ emission. Given that the H$%
_{2}$ emission is coming from this infrared cloud and that it is located on
the opposite side of the cloud from the masers, it is assumed that the
masers are not related to the H$_{2}$ sources. Therefore, this site is
labelled ``not an outflow'' in Table 1.

\subsubsection{G00.70-0.04 (IRAS 17441-2822)}

This maser site is one of several different regions of maser emission
within the rather complex Sag B2 star forming region. The 5 methanol maser spots
in the group that was observed here are offset from the IRAS\ source by ($\Delta$$\alpha$, $\Delta$$\delta$) = (67\farcs7, 83\farcs5). The masers have no associated radio
continuum emission (Walsh et al. 1998), however there are several nearby radio continuum sources. This maser group is only roughly
linear given the relatively large error bars (0\farcs05) in the relative
offsets between individual components and the fact that the distribution
spans only 0\farcs13. It is perhaps not surprising, therefore, that the was
no detected H$_{2}$ emission at this location (Figure 24). The masers also
do not appear to be associated with any near-infrared continuum source
either.

\subsubsection{G10.47+0.03 (IRAS 18056-1952)}

The site of the well-observed hot core from the ammonia observations of
Cesaroni et al. (1994), the masers here are coincident with the water masers
and the compact radio continuum sources `A' and `B' of Hofner \& Churchwell (1996).
However, Figure 25 shows that there is no near-infrared component to these
radio continuum and ammonia sources. None of the bright sources
to the west of the maser location are see at optical wavelengths in the
Digital Sky Survey. However, not only is there no detectable near-infrared component
to G10.47+0.03, there are also no signs of H$_{2}$ emission in the field. Also present in this field is the maser site G10.48+0.03 lying $\thicksim$35$\arcsec$ north of G10.47+0.03 and consisting of 3 maser spots (Figure 25). 

\subsubsection{G11.50-1.49 (IRAS 18134-1942)}

There are 12 methanol maser spots linearly distributed at this site, however
there is no detectable radio continuum emission (Walsh et al. 1998). Figure
26 shows that the maser site is not directly coincident with any bright
near-infrared source, but is located in an large (40$\arcsec\times 40%
\arcsec$) area of extended near-infrared emission. It is not clear if this
extended near-infrared emission is a star forming cloud (in which case the
emission would be due to UV\ fluorescence) or if it is emission from
outflow shocks. There are two bright sources of H$_{2}$ emission present and
both are very close to lying perpendicular to the position angle of the
methanol maser distribution. This is only one of two sources (the other
being G331.28-0.19) in the survey that have H$_{2}$ emission perpendicular to
the maser position angle which are also likely outflow candidates.

\section{Discussion}

\subsection{The case against linearly distributed methanol masers delineating disks}

The first important result from these observations is the fact that the H$_2$ emission is found to be perpendicular to the methanol maser distribution angle in only 2 of the 28 sources observed (7\%). This result was contrary to what was expected. An even more surprising result was that, of the sites where H$_2$ emission was detected, the emission did not appear to be randomly distributed throughout the field of view. Excluding the ``no detections'' and the H$_2$ sites determined to be radiatively excited (i.e. ``not outflow''), Table 1 shows that the great majority (12/15, 80\%) of these maser sites have H$_2$ emission that is found to be predominantly parallel to the linear distributions of methanol masers. However, the exact nature of the H$_2$ emission observed in this survey is not known. It is possible that all the sources of H$_2$ emission on each field are unrelated to the masers and, in the case of multiple H$_2$ sources on the field, unrelated to each other.  Sources may simply be radiatively excited regions of H$_2$ emission. They may also be shock signatures from stellar sources other than the stellar source exciting the masers. However, while any \emph{one} source of H$_2$ observed may be unrelated to the source exciting the masers, the fact that the H$_2$ emission is preferentially found to be parallel to the maser distributions suggests a general physical relationship between the H$_2$ observed and the masers. Therefore, since there appears to be a link between the methanol masers and the H$_2$ emission, and since the general result of the survey is that the the H$_2$ is situated predominantly parallel to the maser distributions, then these results are contrary to what is expected if the disk hypothesis of methanol masers was true.

The near-stellar environment of an O or early B type star is highly caustic, and it is not expected that circumstellar disks can survive long after accretion stops feeding them. There are also no observations of massive stars with debris disks at later stages of evolution. Therefore, these early accretion phases are most likely the only times when massive stars will have disks and their accompanied outflows. It is therefore surprising that 46\% of the maser sites surveyed do not display any evidence of outflow from H$_2$ emission. The detection of H$_2$ in 8 out of 8 low mass YSOs with known CO outflows (Davis \& Eisl\"{o}ffel 1995), and the detection of H$_2$ in 7 out of 7 high mass YSOs with known CO outflows (Kumar et al. 2002) demonstrates the seemingly strong correlation between CO and H$_2$ emission in outflows from YSOs of all masses. Therefore, the high rate of non-detection of H$_2$ from these sources is an additional result contrary to the hypothesis that methanol masers exist in circumstellar disks around massive stars. 

A possible reason for the non-detection of H$_2$ emission may be the high obscuration towards these massive stellar sources. The earliest phases of massive stellar birth are indeed heavily embedded. However, since 6 of the 10 maser sites where no H$_2$ was detected have a near-infrared continuum source present at 2.14 $\micron$, extinction of the 2.12 $\micron$ H$_2$ photons towards these sites cannot play a large factor in explaining the lack of H$_2$ detected. Another possible reason is that the outflows from these sources may be at the wrong velocity to excite the H$_2$ to emit collisionally. H$_2$ is excited in molecular shocks with velocities from 10 to 50 km s$^{-1}$ (Smith 1994). In the survey by Shepherd \& Churchwell (1996), out of 94 high mass stars, 80\% had CO gas components with velocities in the range of 15 to 45 km s$^{-1}$, which corresponds to the collisional excitation range of H$_2$. Therefore, one might again expect to find a strong correlation between the CO outflows and H$_2$ emission from high mass YSOs similar to what is seen from low-mass mass YSOs. However, it should be noted that massive YSOs have been observed with outflow velocity components ranging 2 orders of magnitude from $\lesssim15$ to more than 3000 km s$^{-1}$ (i.e., Shepherd \& Churchwell 1996; Chlebowski \& Garmany 1991). Once the shock speeds of the outflows begin to exceed 40 km s$^{-1}$, H$_2$ starts to become collisionally dissociated (Pineau des For\^{e}ts et al. 2001).        

\subsection{Methanol masers as tracers of outflow?}

Since there is little alternate wavelength information on outflows for these sources, it begs the question: Are these H$_2$ sources outflow signatures? The non-random distribution of H$_2$ sources with respect to the methanol masers suggests a physical link between the two, however the general appearance of the H$_2$ emission in these fields does not look very much like the collimated structures of Herbig-Haro objects and H$_2$ emission seen coming from lower mass young stars (i.e. Davis \& Eisl\"{o}ffel 1995, Eisl\"{o}ffel 2000). The exception is G320.23-0.28, which has H$_2$ distributed in a bipolar morphology with respect to the maser location, and the western ``lobe'' of H$_2$ appears cometary shaped. This H$_2$ emission is exactly parallel with the maser distribution.

There exists some other evidence that linearly distributed methanol masers have outflows parallel to their position angles. The observations of Walsh, Lee \& Burton (2002) of G323.74-0.26 also show extended shock excited H$_{2}$ emission apparently coming from the region of two methanol maser groups. The more pronounced of these two maser groups, G323.740-0.263, is linearly distributed with a position angle consistent with being parallel to the distribution of H$_2$ emission. Furuya et al. (2002) have observations of G24.78+0.08 that show $^{12}$CO(1--0) in a bilobed outflow parallel to, and centered on, the linearly distributed methanol masers there. Furthermore, observations of G318.95-0.20 (Minier, Wong \& Burton in prep.) using HCN show a bilobed outflow situated more or less north-south and consistent with being parallel with the methanol maser distribution that the outflow is centered on. 

These other observations, along with the results present in this paper, not only make a good case against linearly distributed methanol masers being associated with circumstellar disks, they hint at the possibility that the masers are instead more directly associated with the outflows from these sources. This can be argued simply on the grounds that the outflow tracers seem to be preferentially closer to parallel with the methanol maser distribution angles. Furthermore, it would appear (with the possible exception of G313.77-0.86) that the methanol masers are associated with some process near the center of the outflows.

There are several possibilities as to what exactly the methanol masers would trace if they are associated with outflows. Looking to H$_2$O masers, one sees that they are found to be excited in shocks of protostellar jets in close proximity ($\lesssim100$ AU) to young low-mass stars (i.e., Claussen et al. 1998, Patel et al. 1999). Perhaps methanol masers are excited in protostellar jets from high mass stars in a similar way. In the case of H$_2$O masers tracing the jets of low-mass stars, proper motions of the masers have been measured, for instance, in W3(H$_2$O) by Alcolea et al. (1992), clearly indicating that the masers are indeed associated with a radio jet and are moving along the jet away from the central star. The only proper motion study to date of methanol masers is of G9.62+0.19 (Minier et al. 2000b) which showed proper motions consistent with the masers existing in a wide-angled outflow. Another place that H$_2$O masers are thought to be found is along the working surfaces at the side interfaces of the jet with the surrounding ambient gas (i.e. Strelnitski et al. 2002, and references therein). In this case, one might expect, therefore to see methanol masers preferentially in `X' or `V' shapes, instead of straight lines. Looking at very high resolution ($<$1 mas) structure of methanol maser sites, Minier, Conway \& Booth (2001) find W75N has masers all within a conical structure. Minier et al. (2000b) argues the methanol masers of W75N are at a similar position angle as the outflow from this source and are likely to be outflow related. Furthermore, the methanol masers of G31.28+0.06, S 231, and Mon R2 as seen by Minier et al. (2000a) have a well-defined V-shapes. This V-shape is also present with NGC 7538, however this is the site of perhaps two separate maser sites. 

\subsection{Young massive stars and circumstellar disks}

The idea that methanol masers might exist in circumstellar disks around massive stars has generated a lot of interest in the hypothesis. Several authors have performed research to test this hypothesis (i.e. Phillips et al. 1998; Walsh et al. 1998, 1999, 2001; De Buizer et al. 2000, Minier et al. 2000a; Lee et al. 2001) and all have found their data consistent with the hypothesis. However, though the data was consistent, it was by no means conclusive. The work presented here is the first data that seemingly contradicts the hypothesis that linearly distributed methanol masers are generally found in circumstellar disks. 

However, these observations can be said to be indirect proof that massive stars \emph{do} have circumstellar disks. If the H$_2$ emission seen in these sources is outflow related, then outflow implies a disk exists to feed the central star and collimate the outflow. Though in the case of outflows from massive stars, the opening angles can be larger than 90\degr, and thus do not need much collimation. Such large opening angle could encompass all of the H$_2$ sources in G308.918+0.123 (Figure 2), for instance. These large opening angles may hint to the fact that circumstellar accretion disks may not be necessary for the collimation of outflows in massive stars.

A high mass star with outflow has been observed in detail with no detection of an inner accretion disk (i.e. Feldt et al. 1999). The association between outflow and accretion for massive stars has thus been questioned. There are several possibilities to explain this problem, including: 1) The collimation may be achieved through pressure confinement of a larger-scale equatorial torus or envelope of dense gas (Feldt et al. 1999); 2) Circumstellar disks do exist for a time collimating an outflow and then dissipate leaving remnants of the outflow that are later detected; or 3) High mass stars do not generally form via accretion and so there should be no accretion disks. In this last case, outflows from the massive stellar sources that are observed would have to be explained in some other way. A possibility could come from ``circulation models'' (see Shepherd 2003, and references within) where material in the surrounding molecular cloud is gravitationally attracted to a massive stellar core and is diverted at large radii into mass loss through the magnetic poles. This model can account for the large mass loss observed in outflows from massive stars, as well as the large opening angles with no need for an accretion disk.    

\subsection{Methanol masers in the evolutionary sequence of massive star formation}

Much work has been done to try to fit masers of all species into some sort of evolutionary sequence. Some argue that H$_2$O masers are signposts for the earliest phases of star formation because they seem to be associated with hot molecular cores (HMCs; Cesaroni et al. 1994). These HMCs are seen in ammonia emission and are thought to be an extremely young and heavily embedded stage of massive star formation. Because of extinction, they have no visible or near-infrared emission. In fact, they are so deeply embedded that it has been difficult to observe mid-infrared emission from any of these sources as well (De Buizer 2003; De Buizer et al. 2002b). Furthermore, these massive stars are so young that they have no detectable radio continuum emission either. 

However, still others believe that methanol masers trace these earliest stages of massive star formation (e.g. Walsh et al. 1998). This argument comes from the fact that only 20\% of the sites of methanol maser emission in Walsh et al. (1998) are associated with radio continuum emission. Also these UC H{\scriptsize II} regions are generally compact and small and it is argued that this is evidence for the youth of the UC H{\scriptsize II} stage. However, this can also be said to be at the same time an argument in favor of the idea (De Buizer et al. 2000; Phillips et al. 1998) that methanol masers may be generally associated with less-massive stars, say B1-B4 spectral types, because these types of stars would have no UC H{\scriptsize II} regions or small UC H{\scriptsize II} regions due to the low ionization rates of these stars. Lee et al. (2001) argue that methanol masers are also seen in some known HMCs and therefore add evidence to the idea of methanol masers existing in the early stages of massive star formation. They argue that methanol masers begin to show up sometime during the accretion phase of the star and turn off when the UC H{\scriptsize II} region begins to expand. 

Table 1 lists for each maser location the presence of radio continuum emission and near-infrared emission directly coincident with the masers (within the astrometrical accuracy). Assuming that all of the stellar sources exciting methanol maser emission are more massive than B3 (a point in debate, see for instance Lee et al. 2001; De Buizer et al. 2000; Phillips et al. 1998), then massive stars with no radio continuum emission and no near-infrared emission are thought to be at a much earlier stage of formation than stars with both radio and near-infrared continuum emission. Interestingly, in this survey 8 sources are found to have neither type of emission and 7 are found to have both type of emission. This implies that there is an almost equal number of methanol masers at very early and embedded times as there is when the massive stars have evolved enough to dissipate their placental envelopes and be seen by their radio emission and near-infrared emission. A good example of methanol masers existing at much later stages of stellar formation is NGC 6334F (De Buizer et al. 2002c), where the methanol masers exist near a young massive star (or stars) with an evolved and extended UC H{\scriptsize II} region. 

There is some dispute whether one would see near-infrared emission before radio emission from these embedded massive stellar sources. However, maser sources with near-infrared emission but no radio emission are twice as common in this survey (8 sources) than sources with radio emission but no near-infrared emission (4). Simplified radiative transfer models show that one can get near-infrared emission without radio continuum emission in the earliest embedded stages of O-type stellar formation (Testi et al. 1998). However, while such models may explain the existence of dust reprocessed 2 $\micron$ emission, many near-infrared sources associated with methanol masers have been found to have stellar photospheric emission observed in H (1.65 $\micron$) and even J (1.25 $\micron$) bands (Walsh et al. 1999). It is quite possible that the maser sources with near-infrared emission but no radio emission are stellar sources with a lower, non-ionizing, spectral type than B3 (De Buizer et al. 2000). 

In light of this, and all of the available information on H$_2$O masers, OH masers, and methanol masers that exist, one thing is certain: methanol masers, and masers in general, form in a wide variety of locations and during a variety of stellar phases. It seems gross generalizations about when in the star formation process a certain species of maser turns on and off may be an effort in vain. The only thing for certain about maser formation is that masers will form where the conditions are right for them to form. In reality, the near-stellar region around forming stars (massive stars especially) are diverse and dependent on many environmental variables. This diversity is reflected in the variety of circumstances and locations we observe masers to exist. In general, therefore, masers do not have to exist exclusively at certain stages or with certain processes. What is known for sure is that masers of all kinds are indeed associated with star formation, and that all of these masers appear to exist at many stages of stellar formation. Furthermore, there is evidence now that masers are linked to a variety of processes, such as disks, outflows, and shock fronts. 

\section{Conclusions}

The purpose of these observations was to try to test the hypothesis that
linearly distributed methanol masers exist in, and directly delineate,
circumstellar disks by searching for shock excited H$_2$ outflow signatures perpendicular to the
methanol maser position angle. Surprisingly, H$_2$ emission was found to be perpendicular to the methanol maser distribution angle in only 2 of the 28 sources observed (7\%). Furthermore, of the maser locations observed to have H$_2$ emission, the majority have H$_2$ emission found to be dominantly distributed at a position angle within 45\degr of being \textit{parallel} to the maser position angle. This non-random distribution suggests that there is a physical link between the H$_2$ emission and the masers in general. The fact that H$_2$ is predominantly parallel to the methanol masers further contradicts the circumstellar disk hypothesis of methanol masers.

H$_{2}$ emission can also be due to UV florescence, so for any one source in this survey, follow-up observations (in say, $^{12}$CO or HCN) would be needed to confirm if the H$_2$ is from outflow directly related to the stellar source exciting the masers. Observations by other authors of a small number of linearly distributed methanol maser sources lend agreement to the idea that the H$_2$ from these sources may indeed be shock excited in outflows parallel to the methanol maser distributions. A likely explanation for all of this is that at least some linearly distributed methanol masers may be directly associated with outflows. The methanol masers appear to be located coincident with a stellar source at the center of the outflows in most cases. Perhaps the masers trace the jets or outflow
surfaces near the central (proto-)stellar source. The overall morphology and bipolar nature of the H$_2$ emission from G320.23-0.28 makes it the best candidate from this survey for further observations in disproving the circumstellar disk hypothesis and establishing the possible link between methanol masers and outflow. 

The large number (13/28) of non-detections of H$_2$ from these sites (including sites with H$_2$ but unrelated to outflow), also adds doubt to the general presence of circumstellar accretion disks around young massive stars since the disks are thought to be directly responsible for collimating the outflows. Models have been developed that can describe the outflows that are observed from massive stars without the need for accretion disks. The next generation of sub-millimeter and millimeter telescopes (ALMA and SMA) will be needed to conclusively prove whether or not massive stars have, as a general property, circumstellar disks when they form. Until then, it may be uncertain if massive stars form via accretion like low mass stars.   

It was pointed out in the discussion that masers of all species will form wherever the conditions are right for them to form. Therefore trying to prove that any maser species is linked to a single process or evolutionary stage of star formation may be futile. Water and hydroxyl masers have been observed to be linked to a variety of processes during the many different early evolutionary stages of stars, and methanol masers are also in reality most likely associated with outflows, disks, and shocks concurrently. Though the sample size is small, a result of this survey is that methanol masers are just as likely to exist in the earliest stages of star formation when the star is very embedded as when the massive star is well into the stages when its UC H{\scriptsize II} regions has developed and expanded. 

\section*{Acknowledgments}

Many thanks are given to J. T. Radomski for his critical reading of the manuscript. I would also like to thank the anonymous referee for several suggestions that improved the quality of the paper. Astrometry was performed in part with the help of Aladin, developed by CDS, Strasbourg, France.

\clearpage

\begin{figure}
\includegraphics[height=215mm]{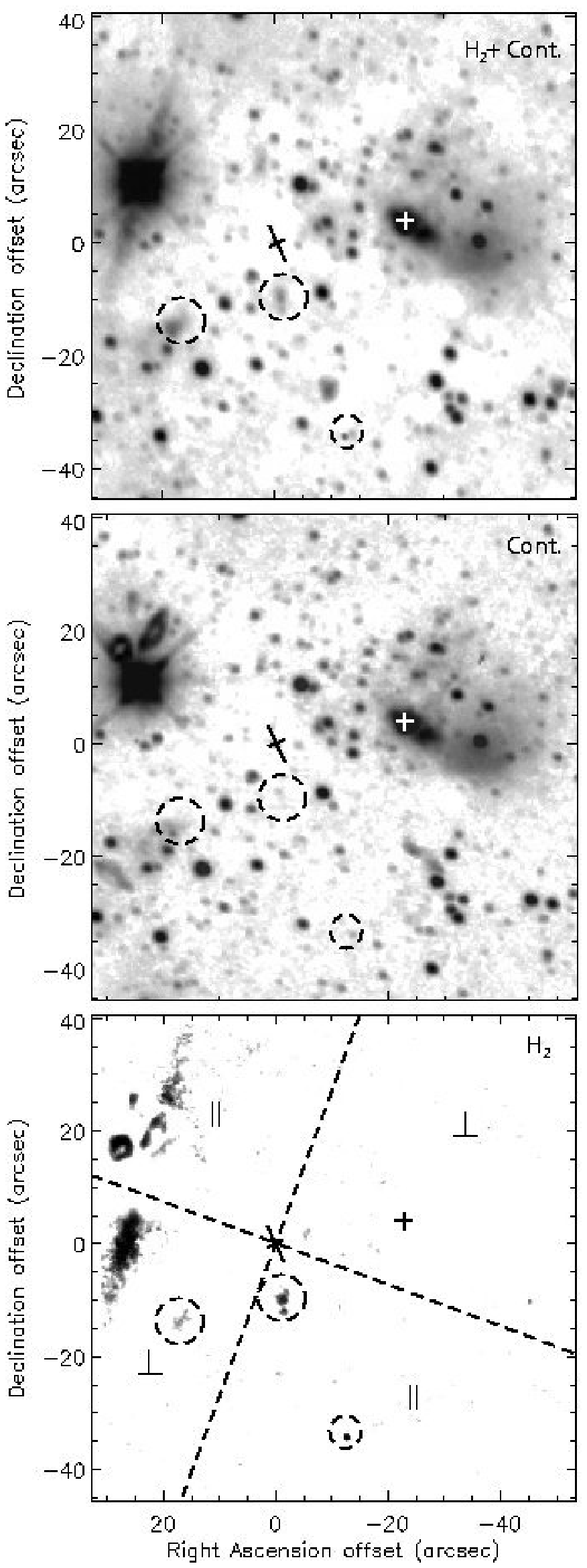}
\caption{G305.21+0.21 (IRAS 13079-6218) H$_2$+continuum, continuum, and residual H$_2$ images. Crosses represent maser group locations, and elongated axes show the position angle of linear maser distributions. Dashed ellipses encompass areas of H$_2$ emission. Dashed lines in the H$_2$ images divide the frame into quadrants parallel and perpendicular to the maser position angle. All emission in the upper left corner of the H$_2$ image is ``noise'' (for definition see \S3) due to the bright stellar source there.}
\end{figure}

\begin{figure}
\includegraphics[height=215mm]{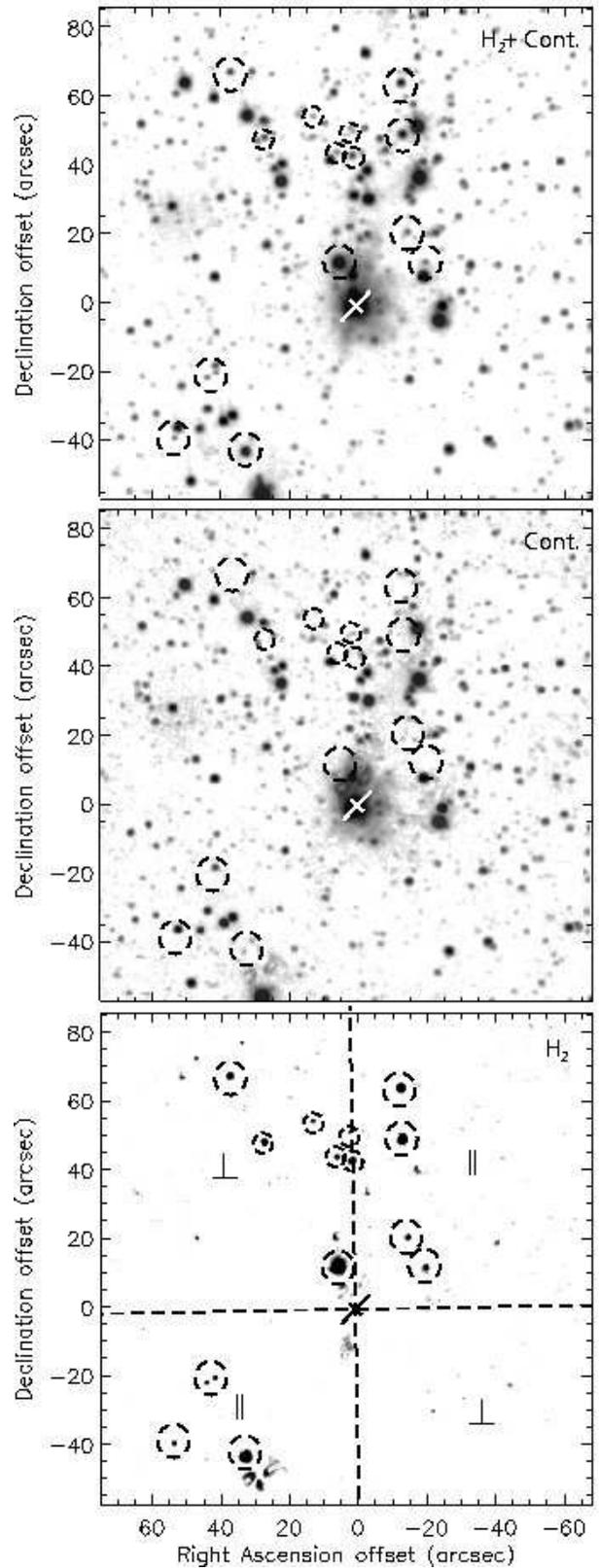}
\caption{G308.918+0.123 (IRAS 13395-6153) H$_2$+continuum, continuum, and residual H$_2$ images. See Figure 1 for symbol descriptions.}
\end{figure}

\clearpage

\begin{figure}
\includegraphics[height=145mm]{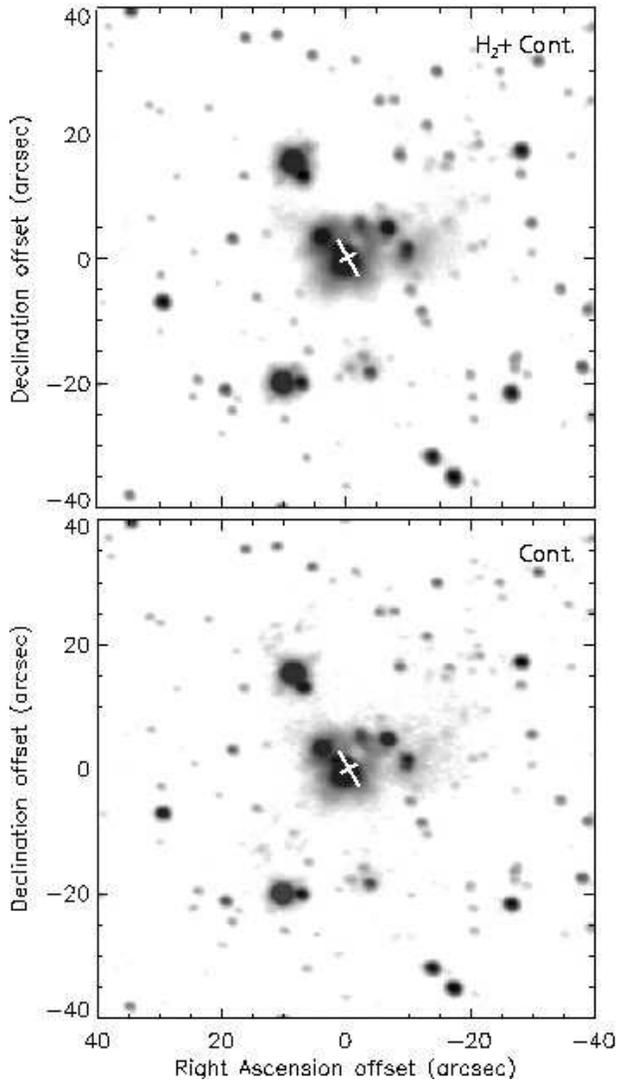}
\caption{G309.92+0.48 (IRAS 13471-6120) H$_2$+continuum, and continuum images. See Figure 1 for symbol descriptions.}
\end{figure}

\begin{figure}
\includegraphics[height=145mm]{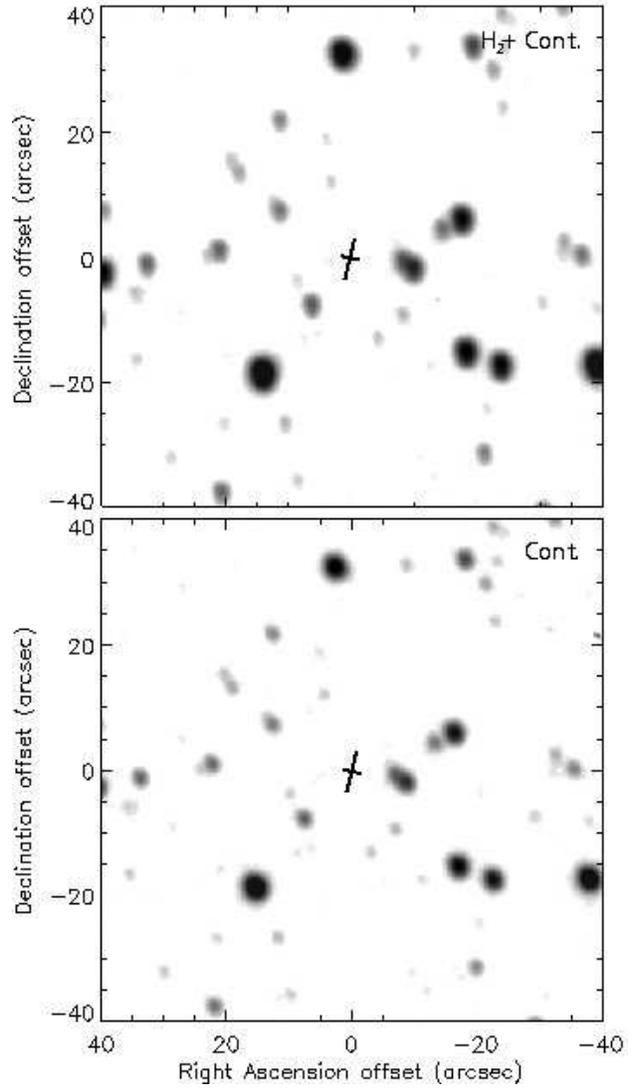}
\caption{G312.11+0.26 (IRAS 14050-6056) H$_2$+continuum and continuum images. See Figure 1 for symbol descriptions.}
\end{figure}

\clearpage

\begin{figure}
\includegraphics[height=215mm]{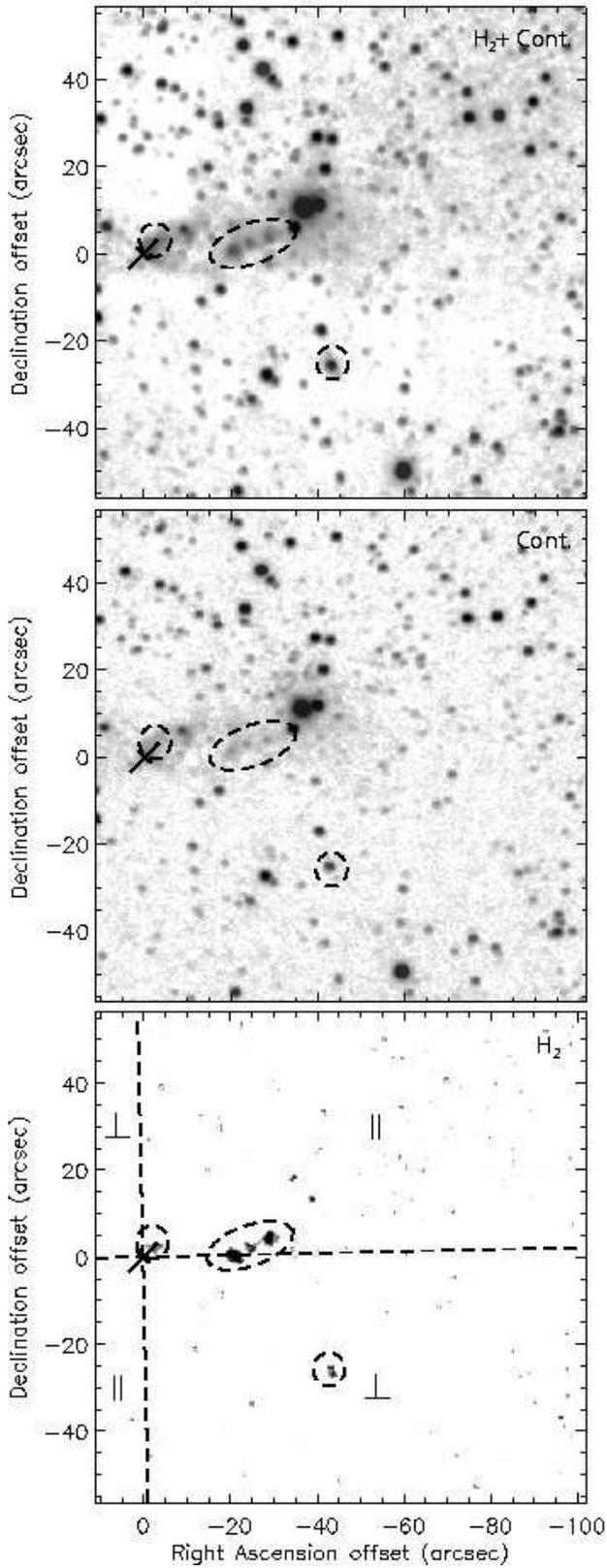}
\caption{G313.77-0.86 (IRAS 14212-6131) H$_2$+continuum, continuum, and residual H$_2$ images. See Figure 1 for symbol descriptions.}
\end{figure}

\begin{figure}
\includegraphics[height=215mm]{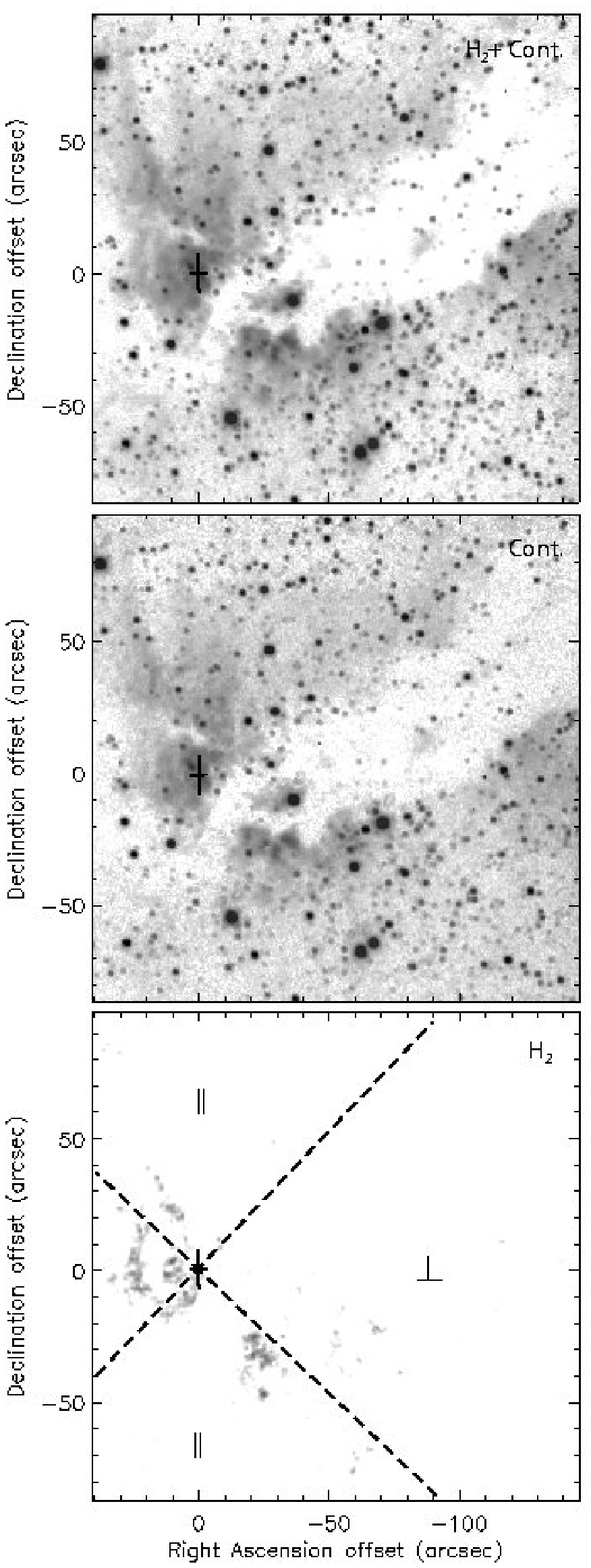}
\caption{G316.81-0.06 (IRAS 14416-5937) H$_2$+continuum, continuum, and residual H$_2$ images. See Figure 1 for symbol descriptions. Because the H$_2$ emission here is not believed to be in outflow, no emission is circled.}
\end{figure}

\clearpage

\begin{figure}
\includegraphics[height=215mm]{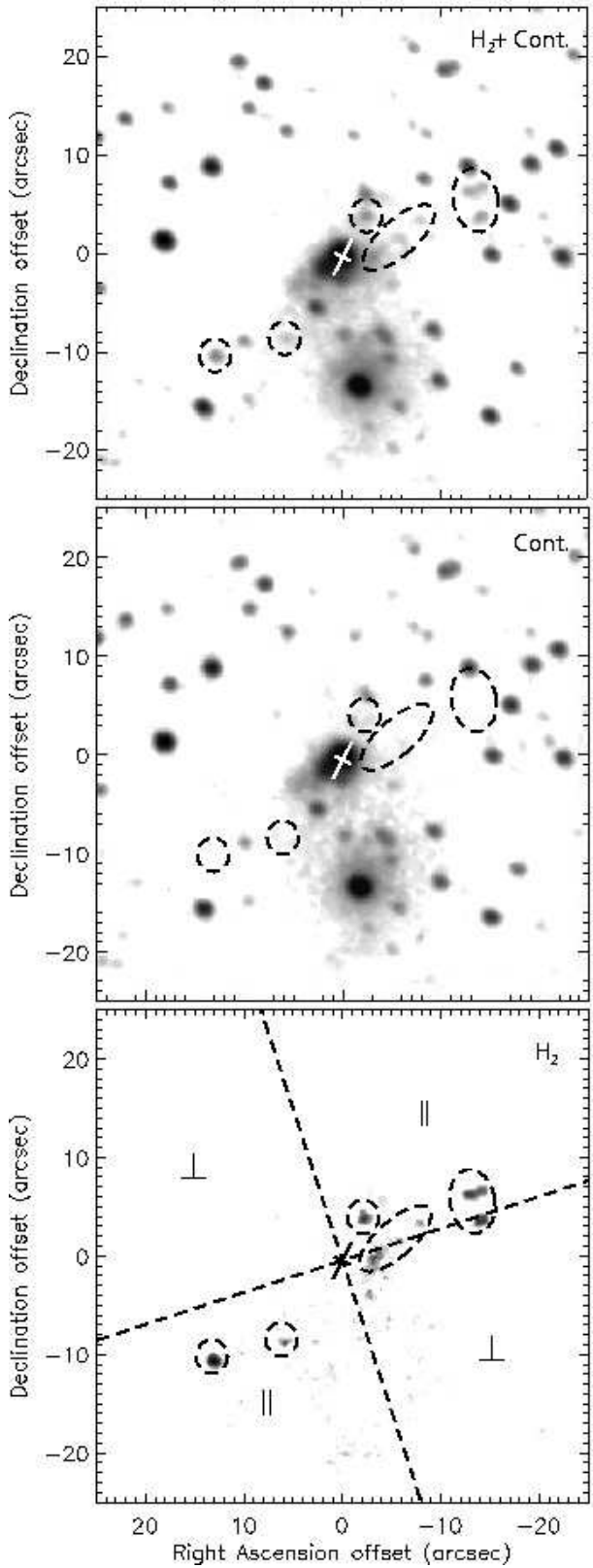}
\caption{G318.95-0.20 H$_2$+continuum, continuum, and residual H$_2$ images. See Figure 1 for symbol descriptions.}
\end{figure}

\begin{figure}
\includegraphics[height=215mm]{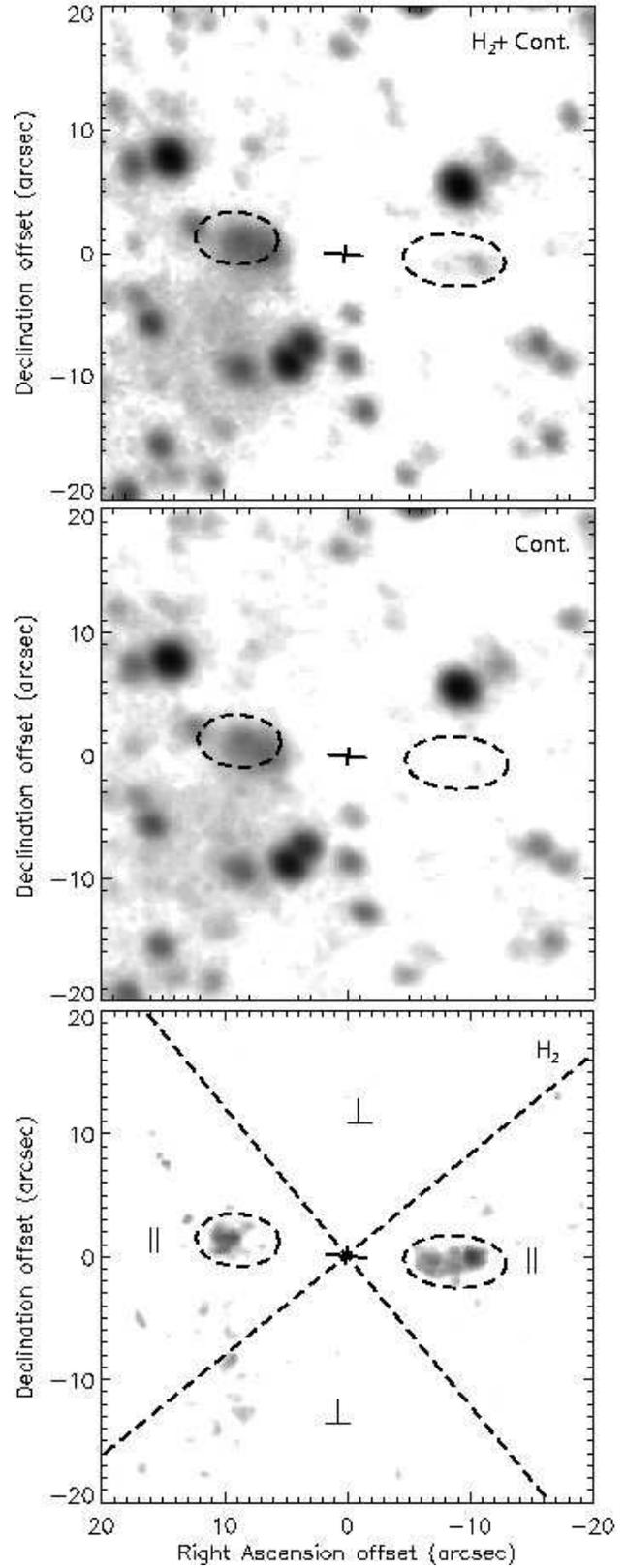}
\caption{G320.23-0.28 (IRAS 15061-5814) H$_2$+continuum, continuum, and residual H$_2$ images. See Figure 1 for symbol descriptions.}
\end{figure}

\clearpage

\begin{figure}
\includegraphics[height=215mm]{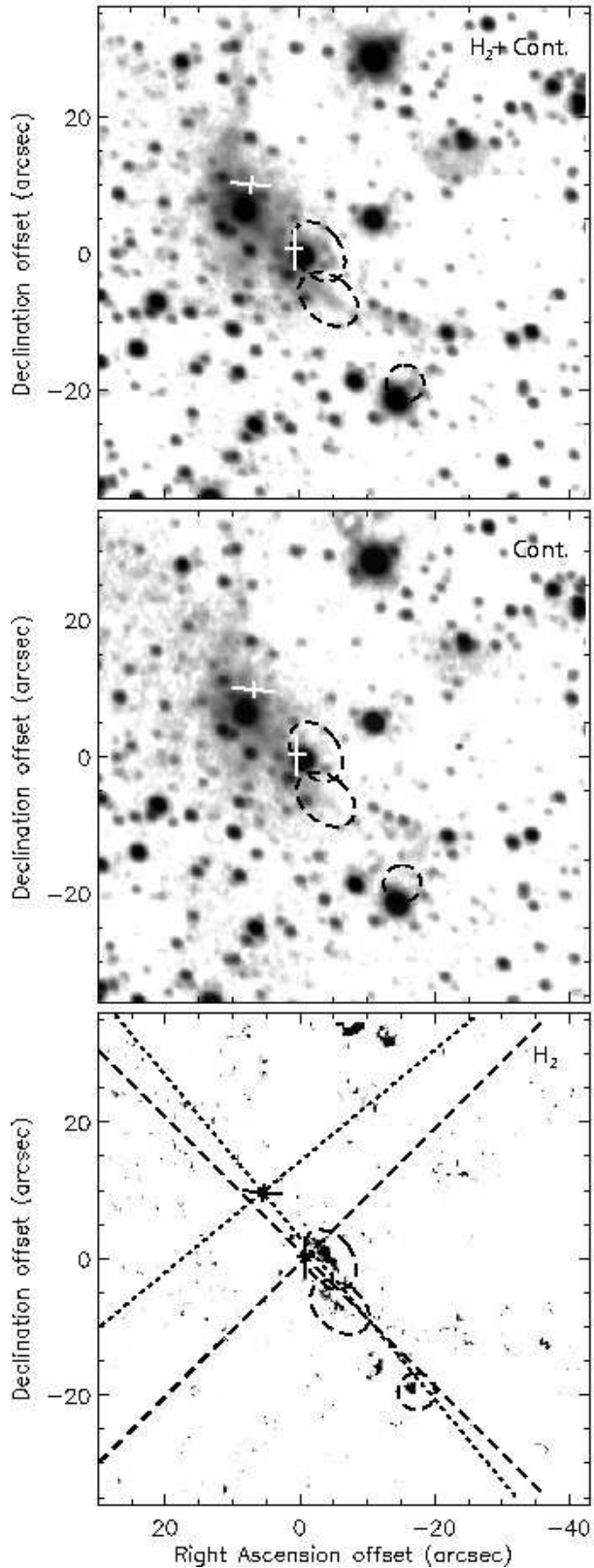}
\caption{G321.031-0.484 (southern cross) and G321.034-0.483 (IRAS 15122-5801) H$_2$+continuum, continuum, and residual H$_2$ images. See Figure 1 for symbol descriptions. The bright residual at the top of the H$_2$ image is noise due to internal reflection.}
\end{figure}

\begin{figure}
\includegraphics[height=145mm]{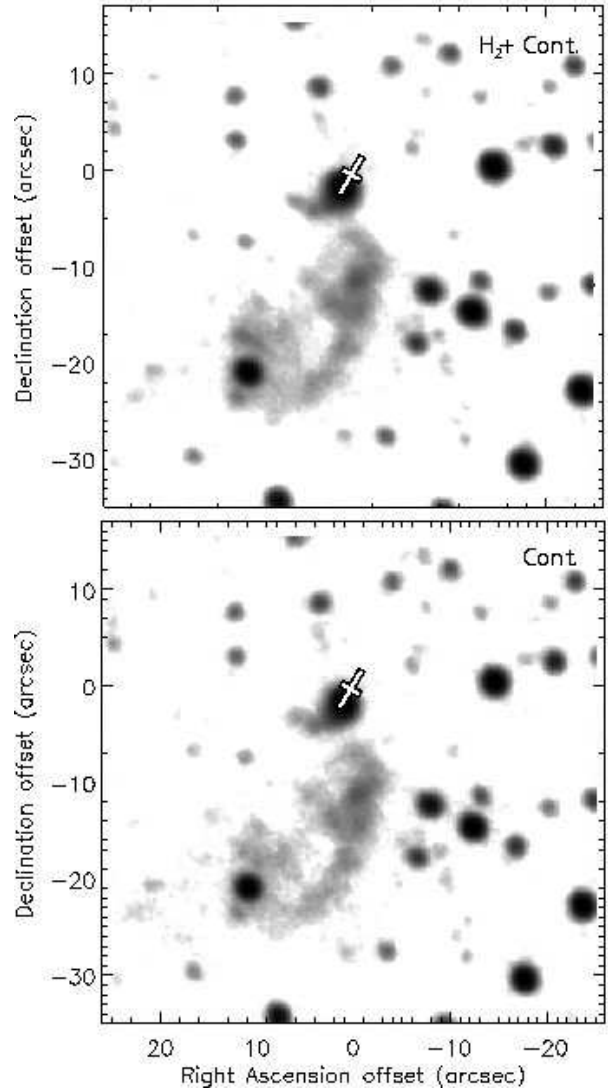}
\caption{G327.120+0.511 (IRAS 15437-5343) H$_2$+continuum and continuum images. See Figure 1 for symbol descriptions.}
\end{figure}

\clearpage

\begin{figure}
\includegraphics[height=145mm]{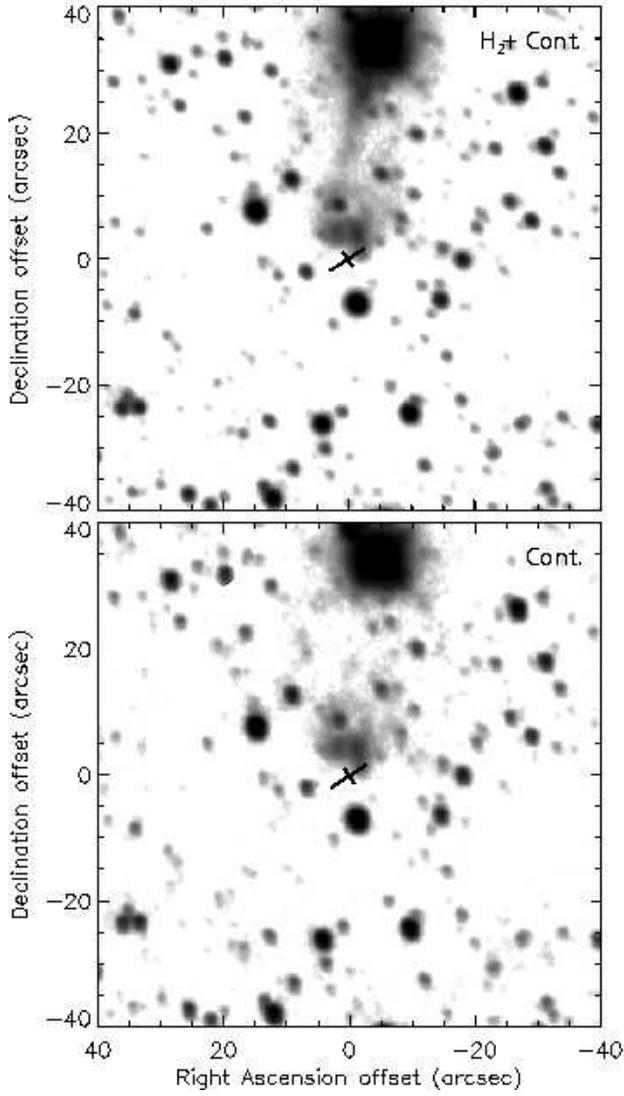}
\caption{G327.402+0.445 (IRAS 15454-5335) H$_2$+continuum and continuum images. See Figure 1 for symbol descriptions. In the H$_2$+continuum image, the emission seen coming from the bright stellar source at the top of the field and extending down towards the maser location is a diffraction spike, and not a real region of H$_2$.}
\end{figure}

\begin{figure}
\includegraphics[height=215mm]{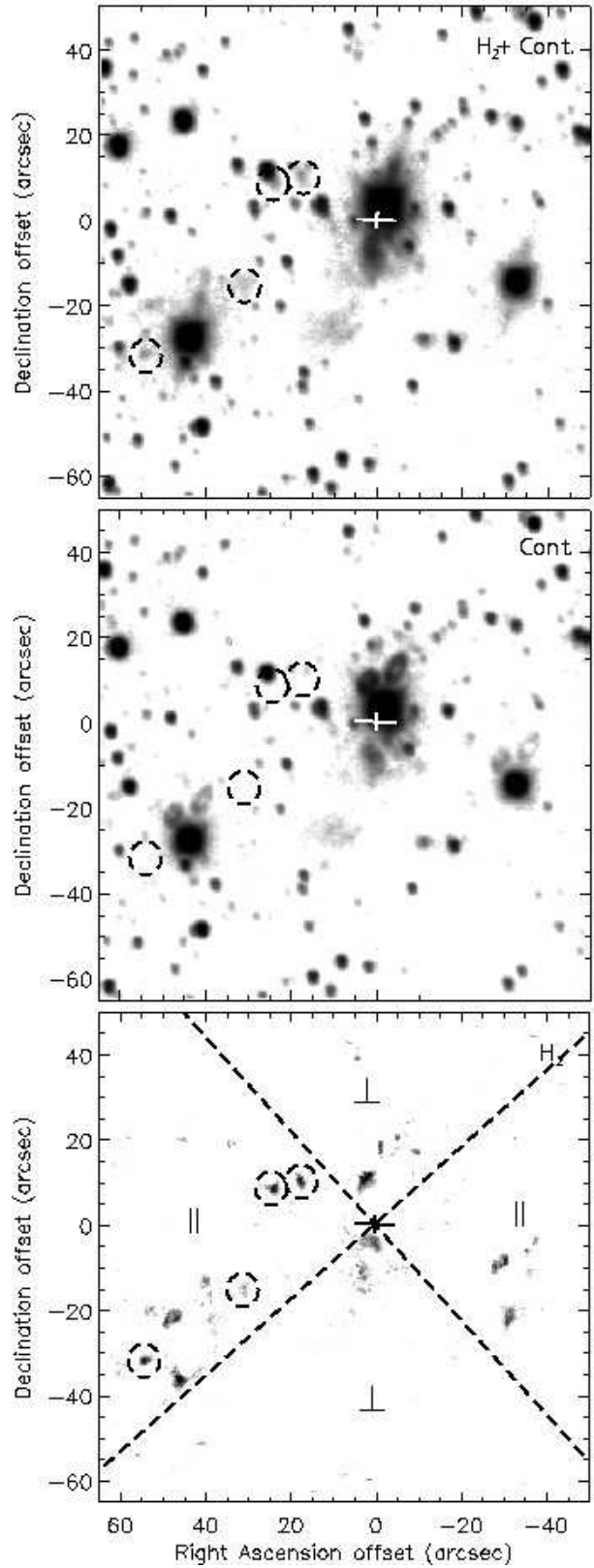}
\caption{G328.81+0.63 (IRAS 15520-5234) H$_2$+continuum, continuum, and residual H$_2$ images. See Figure 1 for symbol descriptions. All three bright sources have internal reflections that can be seen in the H$_2$ frame.}
\end{figure}

\clearpage

\begin{figure}
\includegraphics[height=215mm]{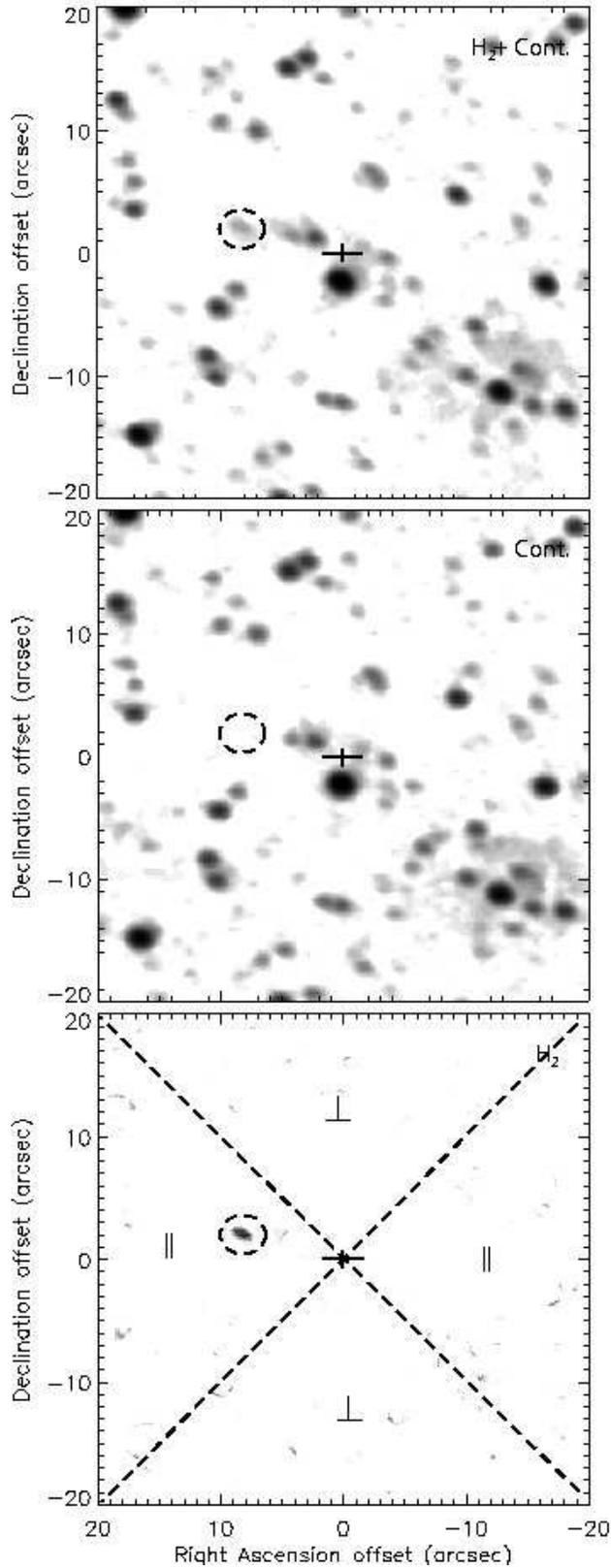}
\caption{G331.132-0.244 (IRAS 16071-5142) H$_2$+continuum, continuum, and residual H$_2$ images. See Figure 1 for symbol descriptions.}
\end{figure}

\begin{figure}
\includegraphics[height=215mm]{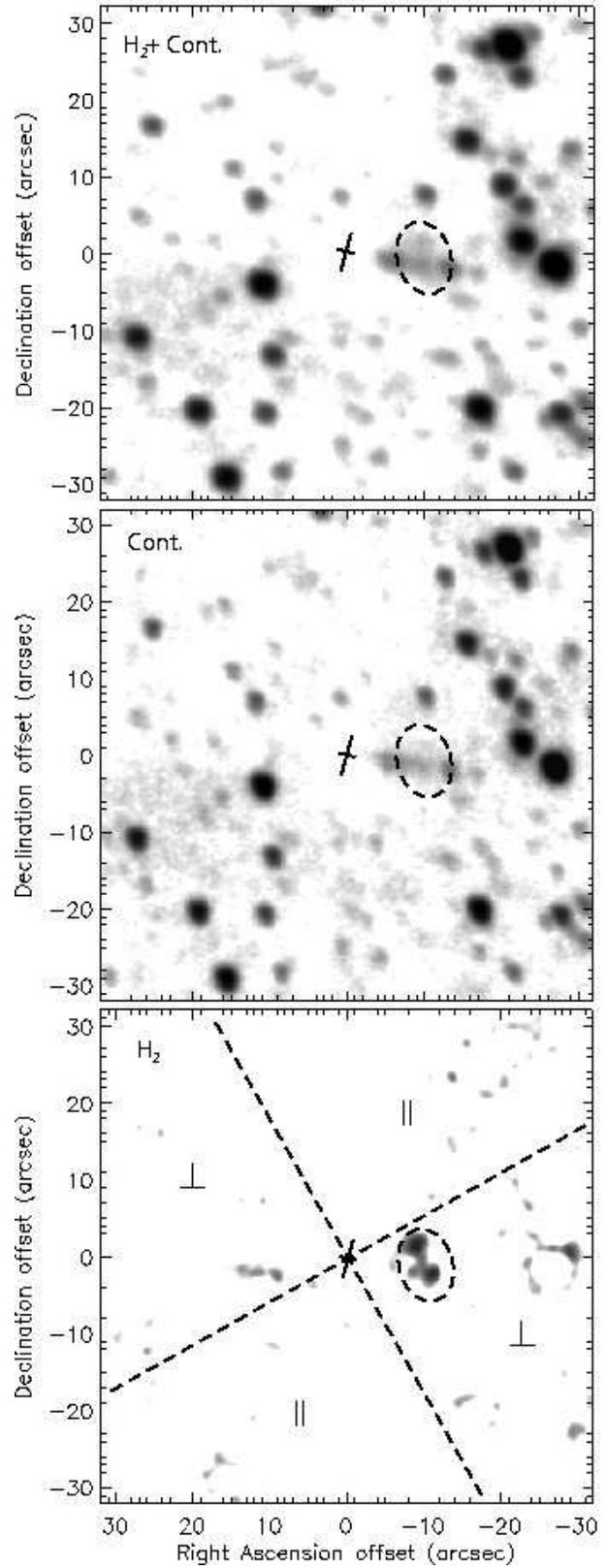}
\caption{G331.28-0.19 (IRAS 16076-5134) H$_2$+continuum, continuum, and residual H$_2$ images. See Figure 1 for symbol descriptions.}
\end{figure}

\clearpage

\begin{figure}
\includegraphics[height=215mm]{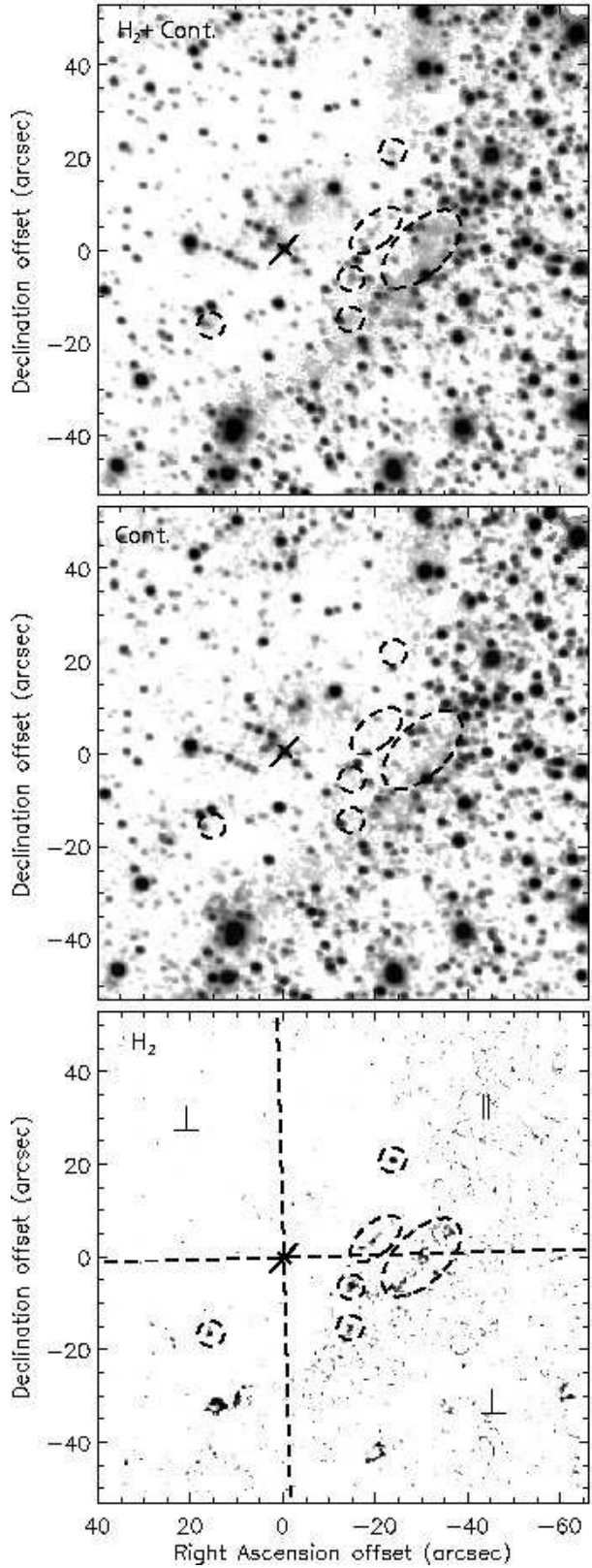}
\caption{G335.789+0.174 H$_2$+continuum, continuum, and residual H$_2$ images. See Figure 1 for symbol descriptions.}
\end{figure}

\begin{figure}
\includegraphics[height=145mm]{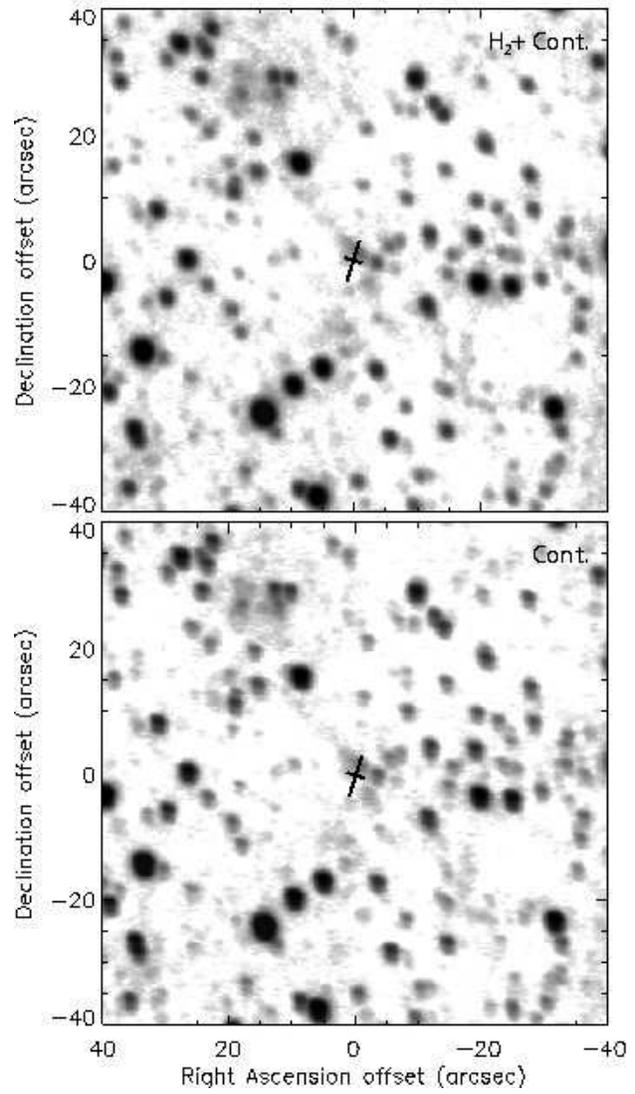}
\caption{G336.43-0.26 (IRAS\ 16306-4758) H$_2$+continuum and continuum images. See Figure 1 for symbol descriptions.}
\end{figure}

\clearpage

\begin{figure}
\includegraphics[height=145mm]{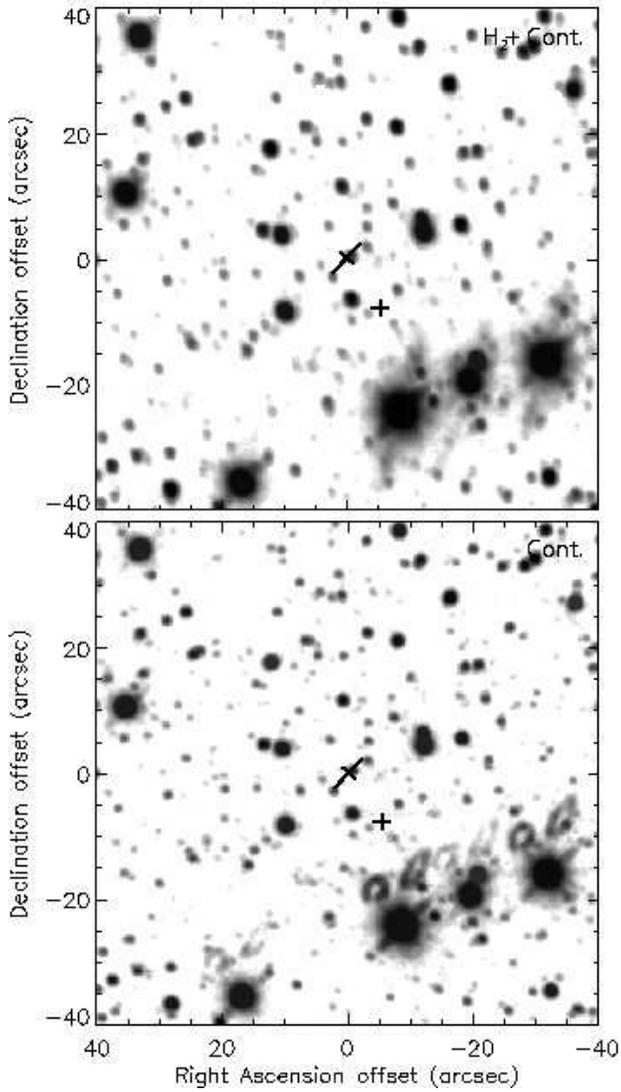}
\caption{G337.705-0.053 (IRAS\ 16348-4654) H$_2$+continuum, and continuum images. See Figure 1 for symbol descriptions. The maser group G337.703-0.053 is also shown as the southern cross.}
\end{figure}

\begin{figure}
\includegraphics[height=215mm]{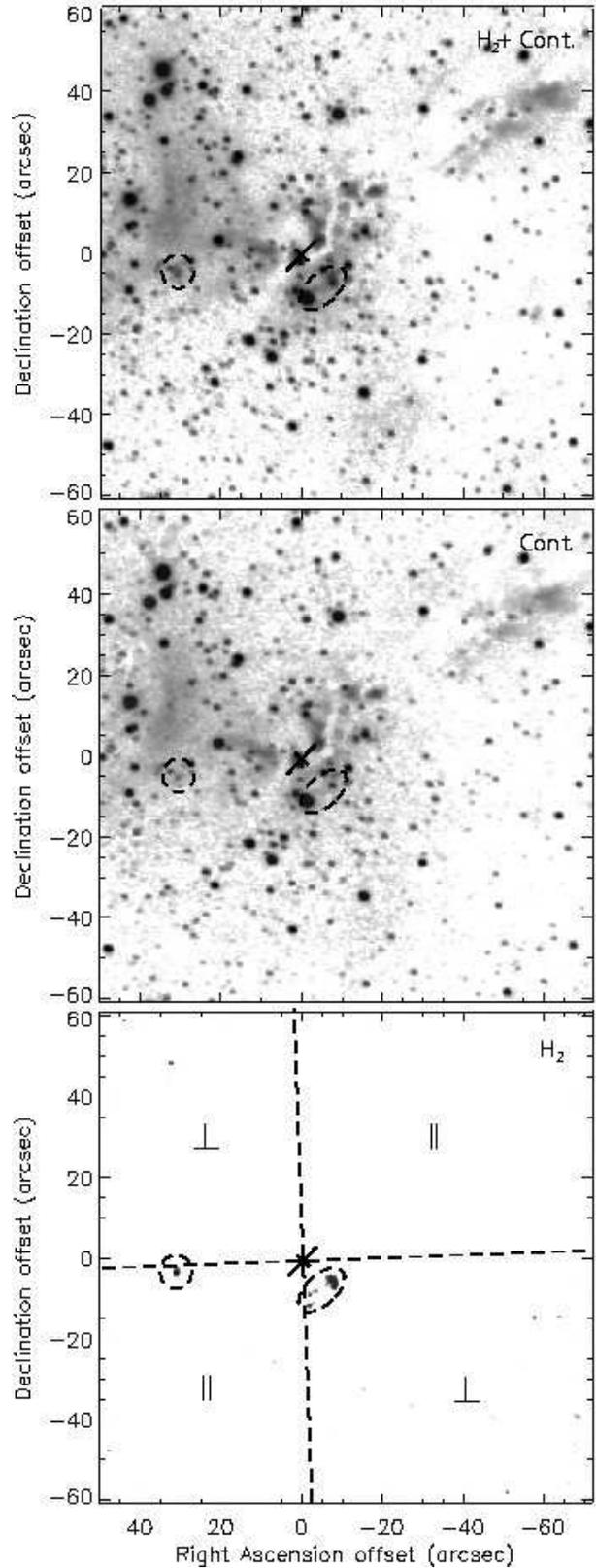}
\caption{G339.88-1.26 (IRAS 16484-4603) H$_2$+continuum, continuum, and residual H$_2$ images. See Figure 1 for symbol descriptions.}
\end{figure}

\clearpage

\begin{figure}
\includegraphics[height=145mm]{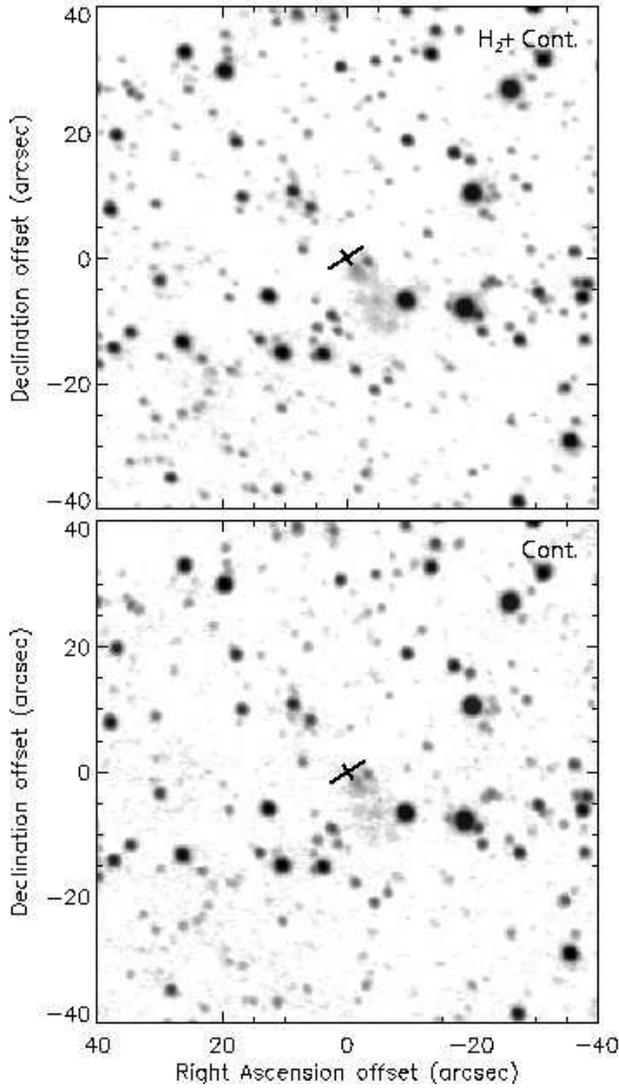}
\caption{G339.95-0.54 (IRAS 16455-4531) H$_2$+continuum and continuum images. See Figure 1 for symbol descriptions.}
\end{figure}

\begin{figure}
\includegraphics[height=145mm]{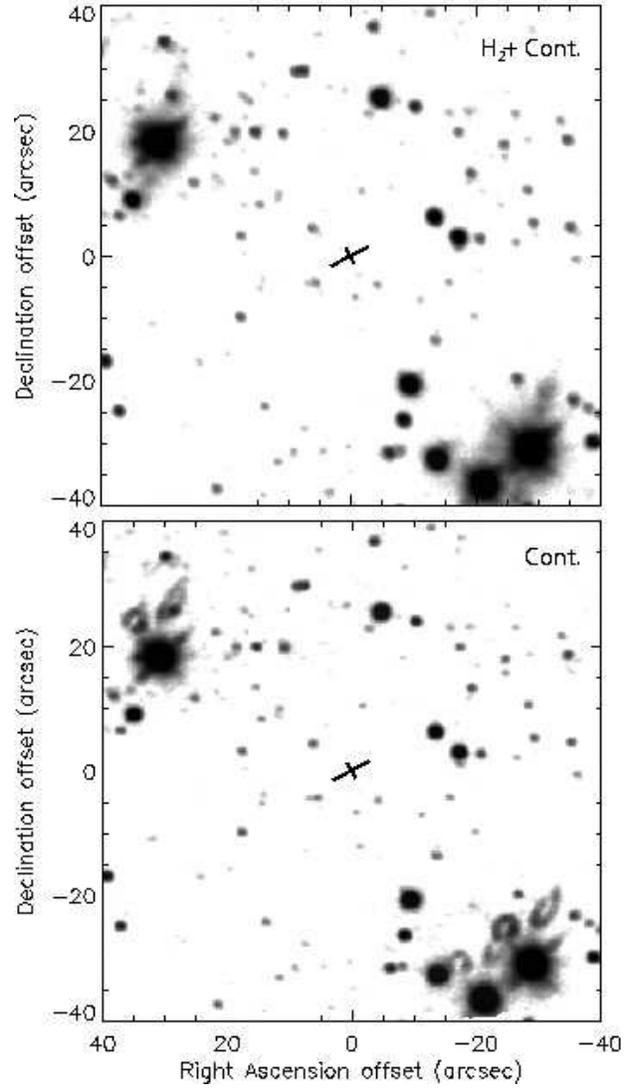}
\caption{G344.23-0.57 (IRAS 17006-4215) H$_2$+continuum and continuum images. See Figure 1 for symbol descriptions.}
\end{figure}

\clearpage

\begin{figure}
\includegraphics[height=215mm]{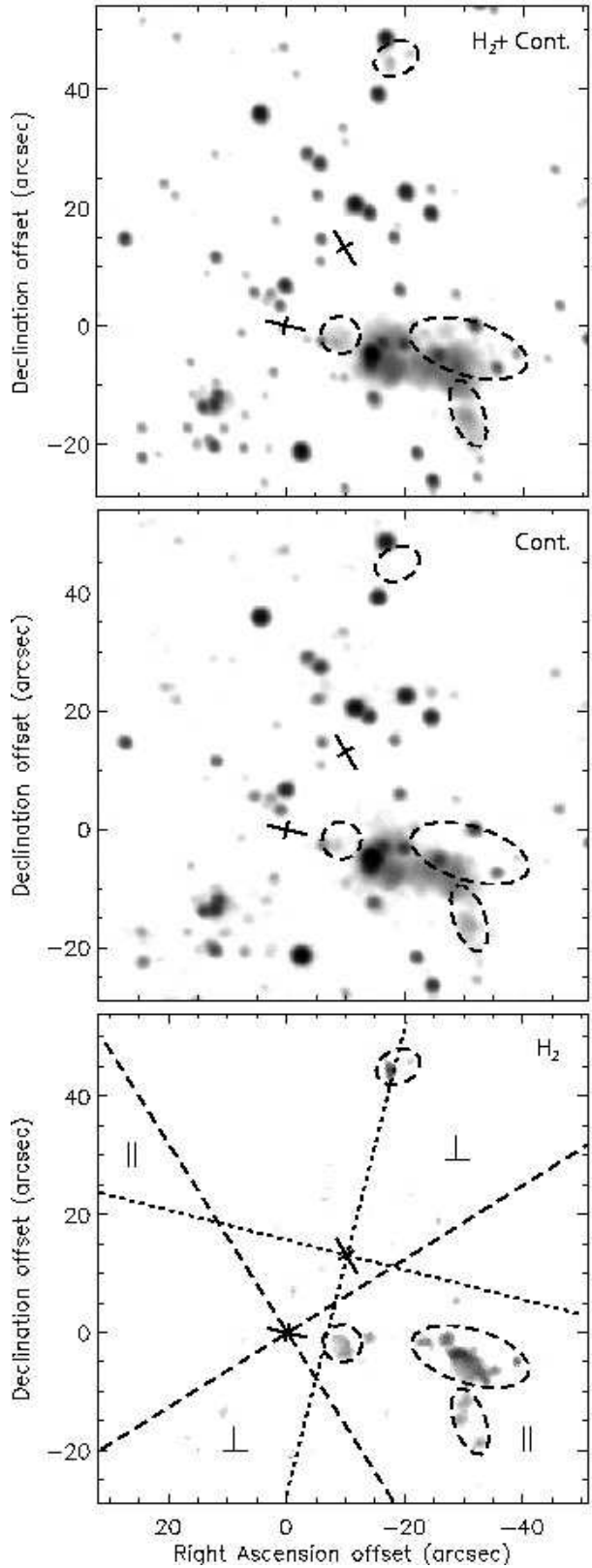}
\caption{G345.01+1.79 (southern cross) and G345.01+1.80 (IRAS 16533-4009) H$_2$+continuum, continuum, and residual H$_2$ images. See Figure 1 for symbol descriptions.}
\end{figure}

\begin{figure}
\includegraphics[height=215mm]{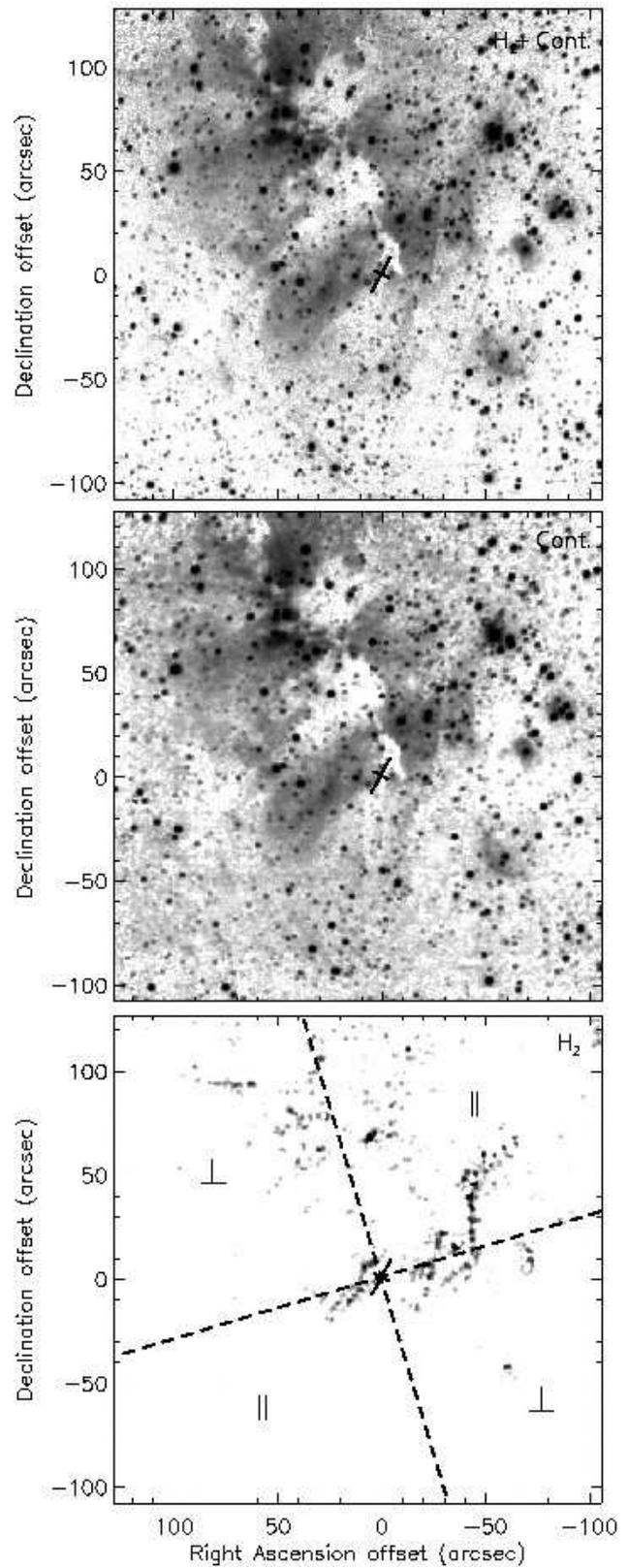}
\caption{G348.71-1.04 (IRAS 17167-3854) H$_2$+continuum, continuum, and residual H$_2$ images. See Figure 1 for symbol descriptions. Because the H$_2$ emission here is not believed to be in outflow, no emission is circled.}
\end{figure}

\clearpage

\begin{figure}
\includegraphics[height=215mm]{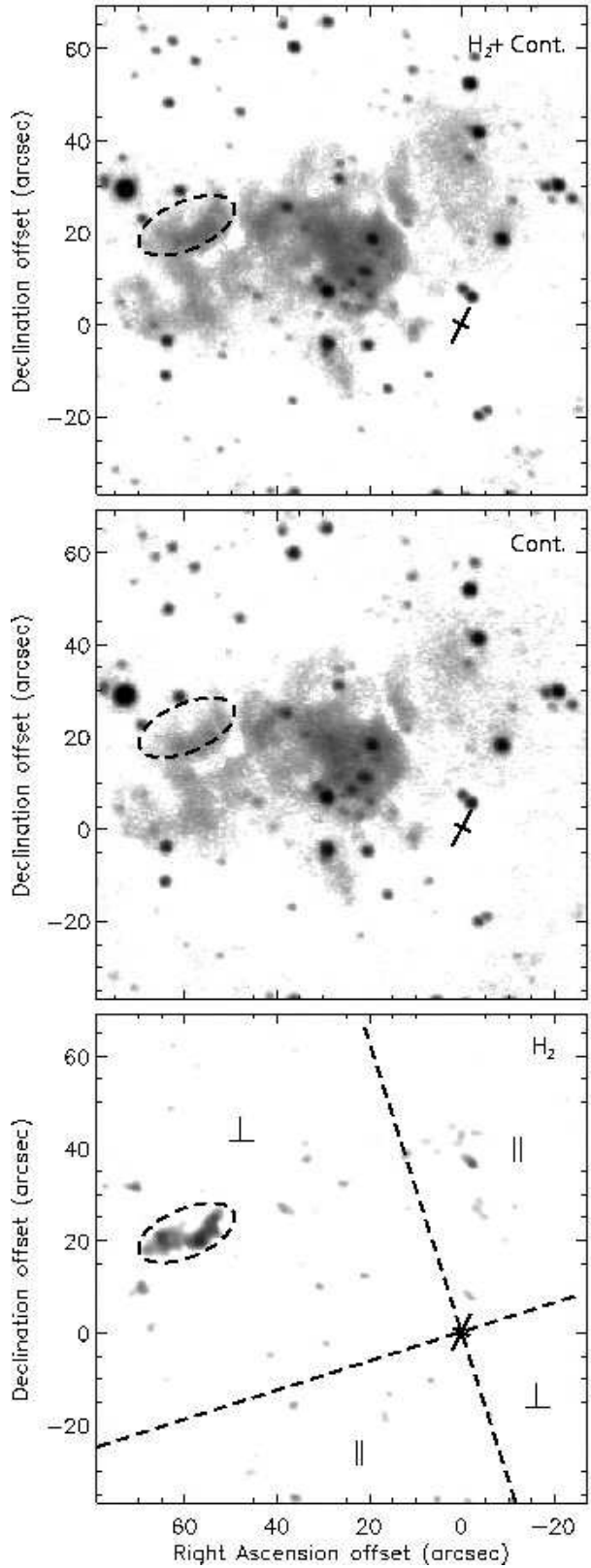}
\caption{G353.410-0.360 (IRAS 17271-3439) H$_2$+continuum, continuum, and residual H$_2$ images. See Figure 1 for symbol descriptions.}
\end{figure}

\begin{figure}
\includegraphics[height=145mm]{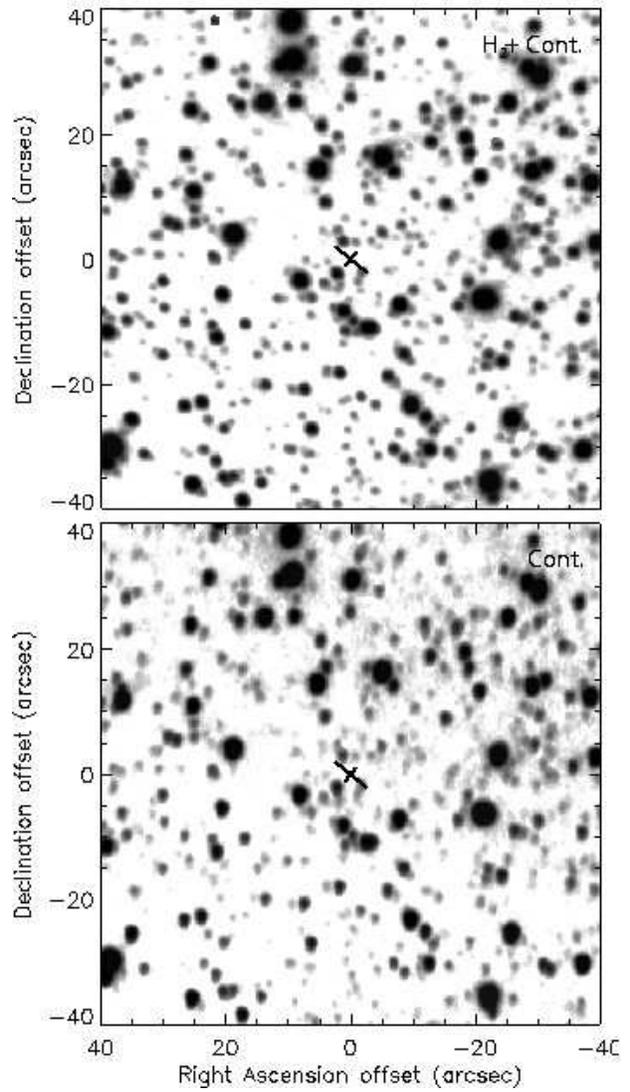}
\caption{G00.70-0.04 (IRAS 17441-2822) H$_2$+continuum and continuum images. See Figure 1 for symbol descriptions.}
\end{figure}

\clearpage

\begin{figure}
\includegraphics[height=145mm]{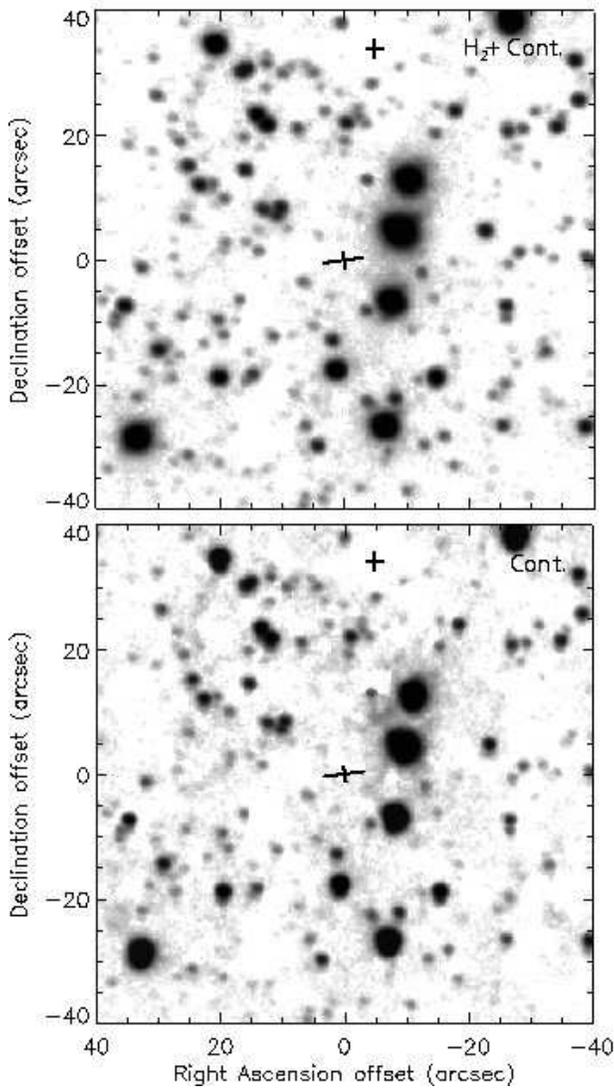}
\caption{G10.47+0.03 (IRAS 18056-1952) H$_2$+continuum and continuum images. See Figure 1 for symbol descriptions. The maser group G10.48+0.03 is also shown as the northern cross.}
\end{figure}

\begin{figure}
\includegraphics[height=215mm]{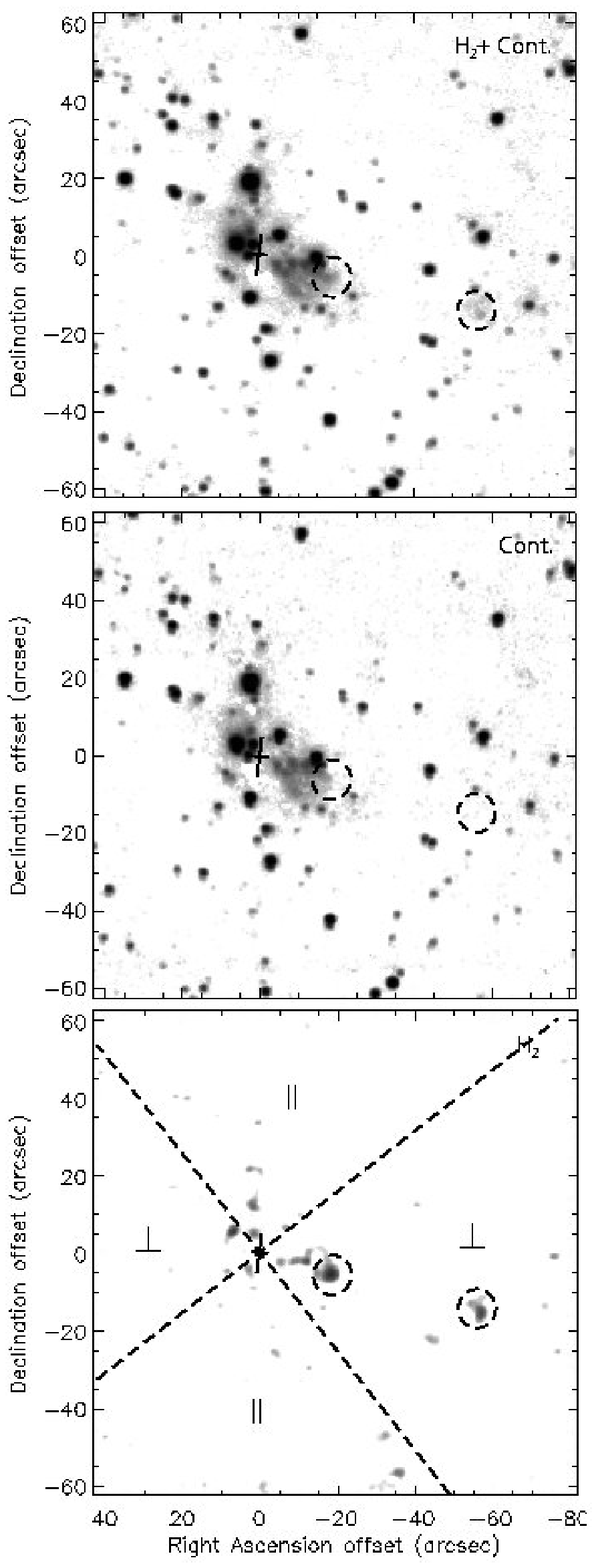}
\caption{G11.50-1.49 (IRAS 18134-1942) H$_2$+continuum, continuum, and residual H$_2$ images. See Figure 1 for symbol descriptions.}
\end{figure}

\label{lastpage}


\begin{thebibliography}{}

\bibitem[\protect\citeauthoryear{Alcolea et al.}{1992}]{A92} Alcolea J., Menten K. M., Moran J. M., Reid M. J., 1992, in Clegg A. W., Nedoluha, G. E., eds, Astrophysical Masers. Springer-Verlag, Heidelberg, p. 225

\bibitem[\protect\citeauthoryear{Bonnell, Bate \& Zinnecker}{Bonnell et al.}{1998}]{BBZ98} Bonnell I. A., Bate M. R., Zinnecker H., 1998, MNRAS, 298, 93

\bibitem[\protect\citeauthoryear{Burton}{1992}]{B92} Burton M. G., 1992, Aust. J. Phys., 45, 463

\bibitem[\protect\citeauthoryear{Caswell}{1998}]{Caswell98} Caswell, J. L., 1998, MNRAS, 297, 215

\bibitem[\protect\citeauthoryear{Caswell et al.}{1995}]{Caswell95} Caswell J. L., Vaile R. A., Ellingsen S. P., Whiteoak J. B., Norris R. P., 1995, MNRAS, 272, 96 

\bibitem[\protect\citeauthoryear{Cesaroni et al.}{1994}]{Ces94} Cesaroni R., Churchwell E., Hofner P., Walmsley C. M., Kurtz S., 1994, A\&A, 288, 903 

\bibitem[\protect\citeauthoryear{Chlebowski \& Garmany}{1991}]{CG91} Chlebowski  T., Garmany, C. D., 1991, ApJ, 368, 241 

\bibitem[\protect\citeauthoryear{Claussen et al.}{1998}]{Clauss98} Claussen M. J., Marvel K. B., Wootten A., Wilking B. A., 1998, ApJ, 507, L79 

\bibitem[\protect\citeauthoryear{Davis \& Eisl\"{o}ffel}{1995}]{DE95} Davis C. J., Eisloeffel J., 1995, A\&A, 300, 851

\bibitem[\protect\citeauthoryear{De Buizer}{2001}]{DeBuizer01} De Buizer J. M., 2001, in Garz\'{o}n F., Eiroa C., de Winter D., Mahoney T. J., eds, ASP Conf. Ser. 219, Disks, Planetesimals, and Planets. ASP, San Francisco, p. 162

\bibitem[\protect\citeauthoryear{De Buizer}{2003}]{DeBuizer03} De Buizer J. M., 2003, in De Buizer, J. M., van der Bliek, N. S., eds, ASP Conf. Ser. 287, Galactic Star Formation Across the Stellar Mass Spectrum. ASP, San Francisco, p. 230

\bibitem[\protect\citeauthoryear{De Buizer, Pina \& Telesco}{De Buizer et al.}{2000}]{DPT00} De Buizer J. M., Pi\~{n}a R. K., Telesco, C. M., 2000, ApJS, 130, 437

\bibitem[\protect\citeauthoryear{De Buizer et al.}{2002a}]{DeBuizer02a} De Buizer J. M., Walsh, A. J. Pi\~{n}a R. K., Phillips C. J., Telesco C. M., 2002a, ApJ, 564, 327

\bibitem[\protect\citeauthoryear{De Buizer et al.}{2002b}]{DeBuizer02b}De Buizer J. M., Watson A. M., Radomski J. T., Pi\~{n}a R. K., Telesco C. M., 2002b, ApJ, 564, L101

\bibitem[\protect\citeauthoryear{De Buizer et al.}{2002c}]{DeBuizer02c}De Buizer J. M., Radomski J. T., Pi\~{n}a R. K., Telesco C. M., 2002c, ApJ, 580, 305

\bibitem[\protect\citeauthoryear{Eisl\"{o}ffel}{2000}]{E00} Eisl\"{o}ffel J., 2000, A\&A, 354, 236

\bibitem[\protect\citeauthoryear{Ellingsen, Norris \& McCulloch}{Ellingsen et al.}{1996}]{ENM96} Ellingsen S. P., Norris R. P., McCulloch P. M., 1996, MNRAS, 279, 101

\bibitem[\protect\citeauthoryear{Feldt et al.}{1999}]{F99} Feldt M., Stecklum B., Henning Th., Launhardt R., Hayward T. L., 1999, A\&A, 346, 243

\bibitem[\protect\citeauthoryear{Forster \& Caswell}{2000}]{FC00} Forster J. R., Caswell J. L., 2000, ApJ, 530, 371

\bibitem[\protect\citeauthoryear{Furuya et al.}{2002}]{F03} Furuya R. S., Cesaroni R., Codella C., Testi L., Bachiller R., Tafalla M., 2002, A\&A, 390, L1 

\bibitem[\protect\citeauthoryear{Hofner \& Churchwell}{1996}]{HC96} Hofner P., Churchwell C., 1996, A\&AS, 120, 283


\bibitem[\protect\citeauthoryear{Kumar, Bachiller \& Davis}{Kumar et al.}{2002}]{KBD02} Kumar M. S. N., Bachiller R., Davis C. J., 2002, ApJ, 576, 313

\bibitem[\protect\citeauthoryear{Lee et al.}{2001}]{Lee01} Lee J.-K., Walsh A. J., Burton M. G., Ashley M. C. B., 2001, MNRAS, 324, 1102 


\bibitem[\protect\citeauthoryear{McKee \& Tan}{2002}]{MT02} McKee C. F., Tan, J. C., 2002, Nature, 416, 59

\bibitem[\protect\citeauthoryear{McKee, Chernoff \& Hollenbach}{McKee et al.}{1982}]{MCH82} McKee C. F., Chernoff D. F., Hollenbach D. J., 1982, in Kessler M. F., Phillips J. P., eds, Proc. of the XVIth ESLAB Symp. 108, Galactic and Extragalactic Infrared Spectroscopy. Reidel, Dordrecht, p. 103  

\bibitem[\protect\citeauthoryear{Minier, Booth \& Conway}{Minier et al.}{2000a}]{MBC00} Minier V., Booth R., Conway J. E., 2000a, A\&A, 362, 1093

\bibitem[\protect\citeauthoryear{Minier, Booth \& Conway}{Minier et al.}{2000b}]{MBECP00} Minier V., Booth R., Ellingsen S. P., Conway J. E., Pestalozzi M. R., 2000b, in Conway J. E., Polatidis A. G., Booth R. S., Pihlstr{\"o}m Y. M., eds, Proc. of the 5th European VLBI Network Symposium, EVN Symposium 2000. Onsala Space Observatory, Onsala, p. 179

\bibitem[\protect\citeauthoryear{Minier, Conway \& Booth}{Minier et al.}{2001}]{MCB01} Minier V., Conway J. E., Booth R., 2001, A\&A 369, 278

\bibitem[\protect\citeauthoryear{Norris et al.}{1993}]{Norris93} Norris R. P., Whiteoak J. B.,
Caswell J. L., Wieringa M. H., Gough R. G., 1993, ApJ, 412, 222

\bibitem[\protect\citeauthoryear{Norris et al.}{1998}]{Norris98} Norris R. P., Byleveld S. E.,
Diamond P. J., Ellingsen S. P., Ferris R. H., Gough R. G., Kesteven M.
J., McCulloch P. M., Phillips C. J., Reynolds J. E., Tzioumis A. K.,
Takahashi Y., Troup E. R., Wellington K. J., 1998, ApJ, 508, 275

\bibitem[\protect\citeauthoryear{Oliva \& Moorwood}{1986}]{OM86} Oliva E., Moorwood A. F. M., 1986, A\&A, 164, 104

\bibitem[\protect\citeauthoryear{Osterloh, Henning \& Launhardt}{Osterloh et al.}{1997}]{OHL97} Osterloh M., Henning Th., Launhardt R., 1997, ApJS, 110, 71

\bibitem[\protect\citeauthoryear{Patel et al.}{1999}]{P99} Patel N. A., Greenhill L. J., Herrnstein J. R., Zhang Q., Moran J. M., Ho P. T. P., Goldsmith P. F., 1999, in Nakamoto T., ed, Star Formation 1999. Nobeyama Radio Observatory, Nagoya, p. 300

\bibitem[\protect\citeauthoryear{Phillips et al.}{1998}]{Phillips98} Phillips C. J., Norris R. P., Ellingsen S. P., McCulloch P. M., 1998, MNRAS, 300, 1131

\bibitem[\protect\citeauthoryear{Pineau des For\^{e}ts et al.}{2001}]{PFASS01} Pineau des For\^{e}ts G., Flower D.R., Aguillon F., Sidis V., Sizun M., 2001, MNRAS, 323, L7

\bibitem[\protect\citeauthoryear{Shepherd}{2003}]{S03} Shepherd D. S., 2003, in De Buizer, J. M., van der Bliek, N. S., eds, ASP Conf. Ser. 287, Galactic Star Formation Across the Stellar Mass Spectrum. ASP, San Francisco, p. 333

\bibitem[\protect\citeauthoryear{Shepherd and Churchwell}{1996}]{SC96} Shepherd D. S., Churchwell E., 1996, ApJ, 457, 267

\bibitem[\protect\citeauthoryear{Shu, Adams \& Lizano}{Shu et al.}{1987}]{SAL87} Shu F. H., Adams F. C., Lizano S., 1987, ARAA, 25, 23

\bibitem[\protect\citeauthoryear{Smith}{1994}]{Smith94} Smith M.D., 1994, MNRAS, 266, 238

\bibitem[\protect\citeauthoryear{Stecklum \& Kaufl}{1998}]{SK98} Stecklum B., Kaufl H., 1998, ESO Press Release PR 08/98

\bibitem[\protect\citeauthoryear{Strelnitski et al.}{2002}]{S02} Strelnitski V., Alexander J., Gezari S., Holder, B. P., Moran J. M., Reid M. J., 2002, astro-ph/0210342

\bibitem[\protect\citeauthoryear{Testi et al.}{1998}]{T98} Testi L., Felli M., Persi P., Roth M., 1998, A\&A, 329, 233 

\bibitem[\protect\citeauthoryear{Vinkovi\'{c} et al.}{2000}]{Vinkovic00} Vinkovi\'{c} D., Miroshnichenko A., Ivezi\'{c} \v{Z}., Elitzur M., 2000, in Sitko M. L., Sprague A. L., and Lynch D. K., eds, ASP Conf. Ser. 196, Thermal Emission Spectroscopy and Analysis of Dust, Disks, and Regoliths. ASP, San Francisco, p. 71

\bibitem[\protect\citeauthoryear{Walsh et al.}{1998}]{Walsh98} Walsh A. J., Burton M. G., Hyland
A. R., Robinson G., 1998, MNRAS, 301, 640

\bibitem[\protect\citeauthoryear{Walsh et al.}{1999}]{Walsh99} Walsh A. J., Burton M. G., Hyland
A. R., Robinson G., 1999, MNRAS, 309, 905

\bibitem[\protect\citeauthoryear{Walsh et al.}{2001}]{Walsh01} Walsh A. J., Bertoldi F., Burton M. G., Nikola T., 2001, MNRAS, 326, 36

\bibitem[\protect\citeauthoryear{Walsh, Lee \& Burton}{Walsh et al.}{2002}]{WLB02} Walsh A. J., Lee J.-K., Burton M. G., 2002, MNRAS, 329, 475

\end{thebibliography}
\end{document}